%% file: main-journal.tex
\documentclass[a4paper, 11pt]{article}

\usepackage{microtype}
\usepackage[utf8]{inputenc}
\usepackage[symbol]{footmisc}
\usepackage{epsfig}
\usepackage{microtype}
\usepackage{graphicx}
\usepackage{enumitem}
\usepackage[linesnumbered,ruled, lined]{algorithm2e}
\usepackage{algpseudocode}
\usepackage{epsfig,enumerate,amsmath,amsfonts,amssymb,amsthm,mathrsfs,ifpdf}
\usepackage{indentfirst,relsize}
\usepackage[numbers]{natbib}
\usepackage{setspace}
\usepackage{enumerate}
\usepackage{latexsym}
\usepackage{stackrel}
\usepackage[all]{xy}
\usepackage[margin = 2.5cm]{geometry}
\usepackage[usenames,dvipsnames]{pstricks}
\usepackage{pst-grad} 
\usepackage{pst-plot} 
\usepackage{xspace}
\usepackage{hyperref}

\usepackage{wrapfig}
\usepackage{tablefootnote}
\newtheorem{theo}{Theorem}[section]
\newtheorem{lem}[theo]{Lemma}

\newtheorem{coro}[theo]{Corollary}

\newtheorem{cl}[theo]{Claim}
\newtheorem{defi}[theo]{Definition}
\newtheorem{rem}{Remark}
\newtheorem{obs}[theo]{Observation}

\input{macro.tex}
\newtheorem{fact}[theo]{Fact}

\renewcommand{\hat}{\widehat}
\renewcommand{\tilde}{\widetilde}
\renewcommand{\Tilde}{\widetilde}
\begin{document}
\title{Efficiently Sampling and Estimating from Substructures using Linear Algebraic Queries}

\author{
Arijit Bishnu\footnote{Indian Statistical Institute, Kolkata, India}
\and 
Arijit Ghosh\footnotemark[1]
\and 
Gopinath Mishra\footnote{University of Warwick, UK}
\and 
Manaswi Paraashar\footnotemark[1]
}

\date{}

\maketitle
\begin{abstract}
\input{abstract.tex}
\end{abstract}
\newpage

\input{intro.tex}

\input{lowerbounds-induced-subgraphs}
\input{bilinear.tex}
\input{sample.tex}
\input{more-results.tex}
\input{conclude1.tex}
\newpage

\bibliographystyle{alpha}
\bibliography{reference}
\newpage

\appendix
\input{conclude}
\input{missingproofs}

\input{probabilityresults}
\end{document}

%% file: macro.tex
\newcommand{\iporacle}[1]{$\mbox{{\sc IP}}_{#1}$\xspace}
\newcommand{\dpd}[2]{\langle {#1}, 
#2\rangle}

\newcommand{\bfe}[3]{$\mbox{{\sc Bfe}}_{#1}(#2,#3)$\xspace}

\newcommand{\sau}[3]{$\mbox{{\sc Sau}}_{#1}(#2,#3)$\xspace}

\newcommand{\onebilinear}{$\mbox{{\sc Bilinear Form Estimation}}$\xspace}
\newcommand{\onebfe}{$\mbox{{\sc Bfe}}$\xspace}
\newcommand{\onealmostuni}{$\mbox{{\sc Sample Almost Uniformly}}$\xspace}
\newcommand{\onesau}{$\mbox{{\sc Sau}}$\xspace}
\newcommand{\indedge}{{\sc Induced Edge Estimation}\xspace}
\newcommand{\indsamp}{{\sc Induced Edge Sampling}\xspace}
\newcommand{\bipedge}{{\sc Induced Bipartite Edge Estimation} \xspace}
\newcommand{\bipsamp}{{\sc Induced Bipartite Edge Sampling} \xspace}

\newcommand{\symmetric}{{\sc Symmetric} \xspace}
\newcommand{\diagonal}{{\sc Diagonal} \xspace}
\newcommand{\trace}{{\sc Trace} \xspace}
\newcommand{\permutation}{{\sc Permutation} \xspace}
\newcommand{\doblestochastic}{{\sc Doubly Stochastic} \xspace}
\newcommand{\allonescolumn}{{\sc All Ones Column} \xspace}
\newcommand{\identicalcolumns}{{\sc Identical Columns} \xspace}

\newcommand{\regr}{{\sc Regr}}

\newcommand{\samplelight}{{\sc Sample-Light}\xspace}
\newcommand{\heavy}{{\sc Sample-Heavy}\xspace}

\newcommand{\ths}{ \tau}
\newcommand{\oneaone}{{\bf{1}}^T A {\bf{1}}}
\newcommand{\bx}{{\bf x}\xspace}
\newcommand{\by}{{\bf y}\xspace}
\newcommand{\vecone}{\bf{1}}
\newcommand{\boldvec}[1]{{\bf #1}\xspace}

\newcommand{\tU}{{\Tilde{U}}}
\newcommand{\tV}{{\Tilde{V}}}
\newcommand{\tTh}{\widetilde{\Theta} \xspace}

\newcommand{\FindIntersection}{\textnormal{FIND-INT}}

\newcommand{\tn}[1]{\textnormal{#1}}

\newcommand{\defproblem}[3]{
  \vspace{1mm}
\noindent\fbox{
  \begin{minipage}{0.96\textwidth}
  \begin{tabular*}{\textwidth}{@{\extracolsep{\fill}}lr} #1 \\ \end{tabular*}
  {\bf{Input:}} #2  \\
  {\bf{Output:}} #3
  \end{minipage}
  }
  \vspace{1mm}
}

\newcommand{\size}[1]{\left| #1 \right|}

\newcommand{\E}{\mathbb{E}}
\newcommand{\remove}[1]{}

\newcommand{\R}{\mathbb{R}}
\newcommand{\N}{\mathbb{N}}

\newcommand{\cM}{\mathcal{M}}

\newcommand{\cL}{\mathcal{L}}
\newcommand{\cA}{\mathcal{A}}

\newcommand{\cE}{\mathcal{E}}

\newcommand{\cP}{\mathcal{P}}

\newcommand{\Oh}{\mathcal{O}}

\newcommand{\tOh}{\widetilde{{\mathcal O}}}

\newcommand{\cH}{\mathcal{H}}

\newcommand{\eps}{\epsilon}
\newcommand{\pr}{\mathbb{P}}

\newcommand{\complain}[1]{\textcolor{red}{#1}}
\newcommand{\comments}[1]{\textcolor{blue}{\bf{#1}}}

%% file: abstract.tex
Given an unknown $n \times n$ matrix $A$ having non-negative entries, the \emph{inner product} ($\mbox{{\sc IP}}$) oracle takes as inputs a specified row (or a column) of $A$ and a vector $v \in \R^{n}$, and returns their inner product. A derivative of {\sc IP} is the {\sc induced
degree} query in an unknown graph $G=(V(G), E(G))$ that takes a vertex $u \in V(G)$ and a subset $S \subseteq V(G)$ as input and reports the number of neighbors of $u$ that are present in $S$. The goal of this paper is to understand the strength of the inner product oracle. Our results in that direction are as follows: (i)
  {\sc IP} oracle can {}{solve} bilinear form estimation, i.e., estimate the value of ${\bf x}^{T}A\bf{y}$ given two vectors ${\bf x},\, {\bf y} \in \R^{n}$ with non-negative entries and can sample almost uniformly entries of a matrix with non-negative entries; (ii) 
  We tackle for the first time weighted edge estimation and weighted sampling of edges that follow as an application to the bilinear form estimation and almost uniform sampling problems, respectively; (iii) 
  {\sc induced degree} query, a derivative of {\sc IP}, can {}{solve} edge estimation and an almost uniform edge sampling in induced subgraphs. To the best of our knowledge, these are the first set of oracle-based query complexity results for induced subgraphs. We show that {\sc IP}/{\sc induced degree} queries over the whole graph can simulate {\sc local} queries in any induced subgraph; (iv) Apart from the above, we also show that {\sc IP} can {}{solve} several problems related to matrix, like testing if the matrix is diagonal, symmetric, doubly stochastic, etc.
The {\sc induced degree} degree query has its roots in the queries that deal with the relation between a vertex and a subset of vertices of a graph as in Ben-Eliezer et al.~[SODA'08] and Nissan~[SODA'21], whereas, the $\mbox{{\sc IP}}$ oracle is in the class of linear algebraic queries used lately in a series of works by Rashtchian et al.~[RANDOM'20], Sun et al.~[ICALP'19], and Shi and Woodruff~[AAAI'19]. The {\sc IP} oracle can be used to estimate the Hamming distance between matrices~[RANDOM'21].

%% file: intro.tex
\section{Introduction}
\label{sec:intro}
\noindent 
Let $G=(V(G),E(G))$ be an unknown graph on $n$ vertices with oracle access, and consider a query on a substructure of $G$.  Suppose we want to know for a vertex $v \in V(G)$, the number of neighbors in a community, as in social networks, represented as a subset $S \subseteq V(G)$. This gives rise to the {\sc induced degree} query oracle \remove{, formally defined as follows. The {\sc induced degree} query} that takes a vertex $u \in V(G)$ and a subset $S \subseteq V$  as input and reports the number of neighbors of $u$ that are present in $S$. On the other hand, let us consider another query, named {\sc inner product} (\iporacle{}) query oracle, the main focus of the paper, with a linear algebraic flavor and show its connection to the {\sc induced degree} query. But let us first deal with the notations used. 


\noindent
{\bf Notations.} In this paper, we denote the set $\{1,\ldots,t\}$ by $[t]$ and $\{0,\ldots,t\}$ by $[[t]]$. For a (directed) graph $G$, $V(G)$ and $E(G)$ denote the vertex set and edge sets of $G$, we will use $V$ and $E$ when the graph is clear from the context. For a vertex $u$, let $d_G(u)$ denote the degree of $u$ in $G$ and $N_G(u)$ denote the set of neighbors of $u$ in $G$. For a subset $S$ of $V(G)$, the subgraph of $G$ induced by $S$ is denoted by $G_{S} = (S, E_{S})$ such that $E_{S} : = \left\{ \{u,v\} \in E(G) \, \mid \, u \in S \mbox{ and } v \in S\right\}$. The {\sc local} queries for a graph $G=(V(G),E(G))$ are: (i) {\sc degree} query: given $u \in V(G)$, the oracle reports the degree of $u$ in $V(G)$; (ii) {\sc neighbor} query: given $u \in V(G)$ and an integer $i$, the oracle reports the $i$-th neighbor of $u$, if it exists; otherwise, the oracle reports $\perp$~\footnote{The ordering of neighbors of the vertices are unknown to the algorithm.}; (iii) {\sc adjacency} query:  given $u, \, v \in V(G)$, the oracle reports whether $\{u,v\} \in E(G)$.

 For a non-empty set $X$ and a given parameter $\eps \in (0,1)$, an \emph{almost uniform} sample of $X$ means each element of $X$ is sampled with probability values that lie in the interval $\left[(1-\eps){1}/{\size{X}},(1+\eps){1}/{\size{X}} \right]$. For a matrix $A$, $A_{ij}$ denotes the element in the $i$-th row and $j$-th column of $A$. $A_{i*}$ and $A_{*j}$ denote the $i$-th row vector and $j$-th column vector of the matrix $A$, respectively. $A \in 
[[\rho]]^{n \times n}$ means $A_{ij} \in [[\rho]]$ for each $i,j \in [n]$, {$\rho \in \N$}. Throughout this paper, the number of rows or columns of a square matrix $A$ is $n$, that will be clear from the context. Vectors are matrices of order n × 1 and will be represented using bold face letters. Without loss of generality, we consider $n$ to be a power of $2$. The $i$-th coordinate of a vector ${\bf x}$ is denoted by 
$x _{i}$. We denote by $\bf{1}$ the vector with all coordinates $1$. Let $\{0,1\}^n$ be the set of $n$-dimensional vectors with entries either $0$ or $1$. For $\bx \in \R^{n}$, ${\bf 1}_{\bx}$ is a vector in $\{0,1\}^{n}$ whose $i$-th coordinate is $1$ if $x_{i} \neq 0$ and $0$ otherwise; $\mathrm{nnz}(\bx) = \size{ i \in [n]: \; x_{i} \neq 0 }$ denotes the number of non-zero components of the vector. 
By $\dpd{\bx}{\by}$, we denote the standard 
inner product of $\bx$ and $\by$, that is, $\dpd{\bx}{\by}=\sum_{i=1}^n x_iy_i$. 
$P$ is a $(1 \pm \eps$)-approximation to $Q$ means $\size{P-Q} \leq \eps \cdot Q$. \remove{
$\tOh(f)-$ $\Oh(f)$ hiding a term polynomial in $\log n$ and $\frac{1}{\eps}$.} \emph{With high probability} means that the probability of success is at least $1-\frac{1}{n^c}$, where $c$ is a positive constant. $\widetilde{\Theta}(\cdot)$ and $\widetilde{\Oh}(\cdot)$ hides a $\mbox{poly}\left(\log n, \frac{1}{\eps}\right)$ term in the upper bound.

\vspace{-4pt}

\subsection{Definition and motivation of {\sc Inner Product} oracle}
\label{ssec:defn}
\noindent
{\bf {\sc inner product} (\iporacle{}) query definition.}
\remove{
{\sc inner product} query oracle, denoted as \iporacle{},  gives access to a matrix $A \in [[\rho]]^{n \times n}$, $\rho \in \N$, whose size is known but the entries are unknown, in the following way. 
}
Let $A \in [[\rho]]^{n \times n}$, $\rho \in \N$, be a matrix whose size is known but the entries are unknown. Now given a row index $i \in [n]$ (or, a column index $j \in [n]$) and a vector ${\bf{v}} \in \R^{n}$, with non-negative entries as input, the {\sc inner product} query to $A$ reports the value of $\dpd{A_{i*}}{\bf{v}}$ ($\dpd{A_{*j}}{\bf{v}}$). \remove{Similarly, given an index $j \in [n]$ for a column and a vector ${\bf{v}} \in S$ as input, the \iporacle{} oracle reports the value of $\dpd{A_{*j}}{\bf{v}}$.} If the input index is for row (column), we refer the corresponding query as row (column) \iporacle{} query. 

Observe that {\sc induced degree} query can be implemented for a graph $G$ by \iporacle{} as a dot product with ${\bf 1}_{S}$ (indicator vector for the set $S$) and the corresponding row of the matrix $A$ that is the $0$-$1$ adjacency matrix of $G$.

\medskip 

\noindent
{{\bf Motivation of the \iporacle{} query.}}
{ 
\noindent The \iporacle{} has both graph theoretic and linear algebraic flavors to it and we will highlight them shortly. It may be mentioned here that \iporacle{} has been already used to estimate the Hamming distance between two matrices~\cite{random21/gopi}. 

From a practical point of view, Rashtchian et al.~\cite{DBLP:conf/approx/RashtchianWZ20} mentions that vector-matrix-vector queries would most likely be
useful in the context of specialized hardware or distributed environments.
Needless to say, the same carries over to \iporacle{}. There are many computer architectures that allow us to compute inner products in one cycle of computation with more parallel processors. Inner product computation can be parallelized using {\em single instruction multiple data} (SIMD) architecture~\cite{compu-arch-book}. Modern day GPU processors use instruction level parallelism. Nvidia GPUs precisely do that by providing a single API call to compute inner products~\cite{sanders10,cuda-dot-nvidia}. There are many such architectures where \iporacle{} query has been given to users directly. Similarly, there are programming languages built on SIMD framework that can compute inner products~\cite{simd-intrinsics}.
}

\subsection{The problems, results and paper organization}
\label{ssec:prob}
The crux of this paper is in showing that \iporacle{} can efficiently solve problems in graph theory that {\sc local} queries and random edge query can not, foremost among them being estimation problems in induced subgraphs, weighted edge estimation, etc. In Section~\ref{sec:induced}, we show this separation with the aid of two lower bounds which give a clear separation in power of \iporacle{} from {\sc local} queries. The problems we consider for showing this separation are edge estimation and sampling problems in induced subgraphs. We show that \iporacle{} can solve these problems using $\tOh\left({\size{S}}/{\sqrt{m_{S}}}\right)$ many \iporacle{} queries to the adjacency matrix of the graph $G$, where $S \subseteq V(G)$ is the set of vertices of interest and {$m_{S}=\size{E_S}$}. \remove{, i.e., the number of edges in the sub-graph of $G$ induced by $S$.} Another crucial takeaway from our work is that \iporacle{} oracle and its derivative, the {\sc induced degree} oracle act like a local query in an induced graph.

\remove{Though we motivated \iporacle{} with graphs, \iporacle{} has a linear algebraic framework.} Our work also involves estimating the bilinear form $\bx^T A \by$ and sampling an element of a matrix almost uniformly using \iporacle{}. \remove{that computes the inner product of a row or column of $A$ with a vector $v \in \R^n$.} Bilinear form estimation has huge importance in numerical linear algebra~(see \cite{FikaMP2014} and the references therein) because of its use in calculating node centrality measures like resolvent subgraph centrality and resolvent subgraph communicability~\cite{BenziKlymko2013,EstradaHigham2010}, Katz score for adapting it to PageRank computing~\cite{BonchiEGGL2012}, etc.  

\noindent
In this paper, we give both upper and lower bounds for several important problems in the context of matrices and graphs when we have \iporacle{} query access to the corresponding matrix and adjacency matrix of the graph, respectively. The main highlights are as follows.

\noindent{\bf Matrix problems.}
The main matrix related  problems considered in this work and defined below are bilinear form estimation and sampling an element of a matrix uniformly at random. 

\defproblem{{\sc Bilinear Form Estimation}$_A(\bx)(\by)$, in short \bfe{A}{\bx}{\by}}{Vectors $\bx \in [[\gamma_1]]^n$, $\by \in [[\gamma_2]]^n$, \iporacle{}  access to matrix $A \in [[\rho]]^{n \times n}$, and  $\eps \in (0,1)$.}{An $(1 \pm \eps)$-approximation to $\bx^TA\by$.}

\defproblem{{\sc Sample-Almost-Uniformly}$_A(\bx)(\by)$, in short \sau{A}{\bx}{\by}}{Vectors $\bx \in [[\gamma_1]]^n$, $\by \in [[\gamma_2]]^n$, \iporacle{}  access to matrix $A \in [[\rho]]^{n \times n}$, and  $\eps \in (0,1)$.}{Report $Z$ satisfying $(1-\eps)\frac{x_iA_{ij}y_j}{\bx^TA\by} \leq \pr(Z=(i,j)) \leq (1+\eps)\frac{x_iA_{ij}y_j}{\bx^TA\by}$.}

\begin{table}
\centering
\begin{tabular}{||c | c |c||} 
 \hline
 {\bf Problem} & {\bf Query complexity} & {\bf Comments}\\
 \hline\hline
 \bfe{A}{\bx}{\by} & $\widetilde{\Theta}\left(\frac{\sqrt{\rho \gamma_1 \gamma_2}\left(\mathrm{nnz}(\bx) + \mathrm{nnz}(\by) \right)}{\sqrt{{\bf x}^{T} A {\bf y}}}\right)$ & Theorem~\ref{theorem-A-not-symmetric-bilinearform-sampling-algo} and~{\ref{theorem-A-not-symmetric-bilinearform-sampling-lower-bound}}\\
\hline
\sau{A}{\bx}{\by} & $\widetilde{\Theta}\left(\frac{\sqrt{\rho \gamma_1 \gamma_2} \left(\mathrm{nnz}(\bx) + \mathrm{nnz}(\by) \right)}{\sqrt{{\bf x}^{T} A {\bf y} }}\right)$ &  Theorem~\ref{theorem-A-not-symmetric-bilinearform-sampling-algo} and~{\ref{theorem-A-not-symmetric-bilinearform-sampling-lower-bound}}\\
 \hline
\end{tabular}
\caption{Query complexities of \bfe{A}{\bx}{\by} and \sau{A}{\bx}{\by}.}\label{table:graphresults1-main}
\end{table} 

\remove{
\begin{table}[h]
\centering
\begin{tabular}{||c | c |c||} 
 \hline
 {\bf Problem} & {\bf Query complexity} & {\bf Comments}\\
 \hline\hline
 \bfe{A}{\bx}{\by} & $\widetilde{\Theta}\left(\frac{\sqrt{\rho \gamma_1 \gamma_2}\left(\mathrm{nnz}(\bx) + \mathrm{nnz}(\by) \right)}{\sqrt{{\bf x}^{T} A {\bf y}}}\right)$ & Theorem~\ref{theorem-A-not-symmetric-bilinearform-sampling-algo} and~{\ref{theorem-A-not-symmetric-bilinearform-sampling-lower-bound}}\\
\hline
\sau{A}{\bx}{\by} & $\widetilde{\Theta}\left(\frac{\sqrt{\rho \gamma_1 \gamma_2} \left(\mathrm{nnz}(\bx) + \mathrm{nnz}(\by) \right)}{\sqrt{{\bf x}^{T} A {\bf y} }}\right)$ &  Theorems~\ref{theorem-A-not-symmetric-bilinearform-sampling-algo} and~{\ref{theorem-A-not-symmetric-bilinearform-sampling-lower-bound}}\\
 \hline \hline
\end{tabular}
\caption{Query complexities of \bfe{A}{\bx}{\by} and \sau{A}{\bx}{\by}. \complain{(Gopi, use hyperref package.)}}
\label{table:graphresults1-main}
\end{table}
}

For the above problems, we give both upper and (almost) tight lower bounds \remove{for bilinear form estimation and sampling an element of a matrix {almost} uniformly at random} (see  Table~\ref{table:graphresults1-main}) in Section~\ref{sec:matrix}. We also discuss weighted edge estimation and weighted sampling of edges as their applications in Appendix~\ref{sec:weighted}. Apart from these, we also discuss (in Appendix~\ref{app:matrix}\remove{ and  Appendix~\ref{sec:stat}}) several important matrix problems using \iporacle{} oracle that were studied using stronger queries like \emph{matrix vector} and \emph{vector matrix vector} queries~\cite{SunWYZ19,DBLP:conf/approx/RashtchianWZ20}.

{\bf Graph problems.} 
Section~\ref{sec:induced} discusses our results for graph problems and establishes tight separation between {\sc induced degree} query and {\sc local} query oracle. To establish the fact that {\sc local} query access (to the entire unknown graph) can not solve problems in induced subgraphs \emph{efficiently}, we prove lower bounds for {\sc local} query access to solve {\sc Edge Estimation} and {\sc Edge Sampling} in induced subgraphs in Section~\ref{sec:induce_lower}. In Section~\ref{sec:induced_upper}, we will discuss that \iporacle{}/{\sc induced degree} for the whole graph can simulate {\sc local} queries in any induced subgraph, and describe its implication in solving problems in induced subgraph.


\section{Inner product oracle vis-a-vis other query oracles}
\label{sec:iporacle}
\noindent
Graph parameter estimation, where the graph can be accessed through query oracles only, has been an active area of research in sub-linear algorithms for a while~\cite{GoldreichR08,EdenLRS15,EdenRS18,RubinsteinSW18}. There are different granularities at which the graph can be accessed -- the query oracle can answer properties about graph that are local or global in nature. By now, the {\sc local} queries have been used for edge~\cite{GoldreichR08}, triangle~\cite{EdenLRS15}, clique estimation~\cite{EdenRS18} and has got a wide acceptance among researchers. 
\remove{
\begin{itemize}
\item[(i)] {\sc degree} query: given $u \in V(G)$, the oracle reports the degree of $u$ in $V(G)$;
\item[(ii)] {\sc neighbor} query: given $u \in V(G)$ \complain{and an integer $i$}, the oracle reports the $i$-th neighbor of $u$, if it exists; otherwise, the oracle reports $\perp$;
\item[(iii)] {\sc adjacency} query:  given $u, \, v \in V(G)$, the oracle reports whether $\{u,v\} \in E(G)$.
\end{itemize}
}
Apart from the local queries, in the last few years, researchers have also used the {\sc random edge} query~\cite{AliakbarpourBGP18,AssadiKK19}, where the oracle returns an edge in the graph $G$ uniformly at random. Notice that the randomness will be over the probability space of all edges, and hence, it is difficult to classify a random edge query as a local query. 
On the other hand, \emph{global queries} come in different forms. Starting with the subset queries~\cite{Stockmeyer83,Stockmeyer85,RonT16}, there have been other queries like {\sc bipartite independent set} query, {\sc independent set} query~\cite{BeameHRRS18}, {\sc gpis} query~\cite{BishnuGKM018,DellLM20}, {\sc cut} query~\cite{RubinsteinSW18}, etc. Linear measurements or queries~\cite{AssadiCK21,AhnGM12}, based on dot product, have been used for different graph problems. 

To this collection of query oracles, we introduce a new oracle called {\sc inner product} (\iporacle{}) oracle that is a natural oracle to consider for linear algebraic and graph problems. Using this oracle, 
we {solve} hitherto unsolved problems ({by an unsolved problem, we mean that no non-trivial algorithm was known before}) with graph theoretic and linear algebraic flavor, like (a) edge estimation in induced sub-graph; (b) bilinear form estimation; (c) sampling entries of matrices with non-negative entries. We also show weighted edge estimation and edge estimation in induced subgraph as applications of bilinear form estimation. Our lower bound result, for {\sc Edge Estimation} in induced subgraph  with only {\sc local} query access, implies that there is a separation between the powers of {\sc local} query and {\sc induced degree} query. \remove{The introduction of a new query oracle model is going to be contentious. We feel the justification for introducing a new query oracle lies in a few issues -- (i) can it solve a problem that has not been considered in the query complexity setting? (ii) can it solve already solved problems (in some other query models) but with appreciably lesser queries? (iii) does there exist a clear separation in the power of this oracle in solving some problems vis-a-vis existing query oracles? 
{\sc cut} queries of Rubinstein et al.~\cite{RubinsteinSW18} is an example of (i) and (iii), where a new query oracle was used to solve a new problem (min-cut finding); our new query oracle also satisfies this condition by solving new natural problems in graphs and matrices. {\sc bis} and {\sc gpis} query oracles~\cite{BeameHRRS18,BhattaBGM19,DellLM20} are examples of stronger global queries that solve edge estimation in graphs~\cite{BeameHRRS18} or hyperedge estimation in hypergraphs~\cite{BhattaBGM19,DellLM20} using polylogarithmic queries as opposed to sub-linear local queries (but much higher than polylogarithmic queries)~\cite{GoldreichR08}. Here, we would like to note that \iporacle{} satisfies not only (i), but also (ii) and (iii).}We will show that our newly introduced {\sc inner product} query oracle can solve problems that can not be solved by the three local queries mentioned even coupled with the random edge query.


\remove{
Moving on to graphs, suppose we want to know for a vertex $v \in V(G)$, the number of neighbors in a community, as in social networks, represented as a subset $S \subseteq V(G)$. This gives rise to the {\sc induced degree} oracle, formally defined as follows. The {\sc induced degree} query takes a vertex $u \in V(G)$ and a subset $S \subseteq V$  as input and reports the number of neighbors of $u$ that are present in $S$. Observe that {\sc induced degree} query can be implemented for a graph $G$ by \iporacle{} as a dot product with ${\bf 1}_{S}$ (indicator vector for the set $S$) and the corresponding row of the matrix $A$ that is the $0$-$1$ adjacency matrix of $G$.} 

Our current survey of the literature (here we do not claim exhaustivity!) shows that a query related to a subgraph was first used in Ben-Eliezer et al.~\cite{Ben-EliezerKKR08}, and named as \emph{group query}, where one asks if there is at least one edge between a vertex and a set of vertices. We found the latest query in this league to be the \emph{demand query} (in bipartite graphs where the vertex set are partitioned into two parts left vertices and right vertices) introduced by Nissan~\cite{abs-1906-04213} -- a demand query accepts a left vertex and an order on the right vertices and returns the first vertex in that order that is a neighbor of the left vertex. One can observe that the group and demand queries are polylogarithmically equivalent. Staying on this line of study related to the relation of a vertex with a subset of vertices, we focus on the {\sc induced degree} query which we feel handles many natural questions.

\remove{Another link that motivates our study is the fact that the {\sc induced degree} query has a natural interpretation as a linear algebraic query supported on an adjacency matrix, and can be generalized to the \iporacle{} query.}
Query oracle based graph algorithms access the graph at different granularities -- this gives rise to a whole gamut of queries with different capacities, ranging from local queries like degree, neighbor, adjacency queries~\cite{Feige06,GoldreichR08} to global queries like independent set based queries~\cite{BeameHRRS18,DellLM20}, random edge queries~\cite{AliakbarpourBGP18}, and others like group~\cite{Ben-EliezerKKR08} and demand queries~\cite{abs-1906-04213}. This rich landscape of queries has unravelled many interesting algorithmic and complexity theoretic results~\cite{Feige06,GoldreichR08,AliakbarpourBGP18, Ben-EliezerKKR08,abs-1906-04213,BeameHRRS18,DellLM20,DBLP:journals/toct/RonT16}. With this in mind, if we turn our focus to the landscape of linear algebraic queries, the most natural query is the \emph{matrix entry} query where one gives an index of the matrix and asks for the value there. Lately, a series of works~\cite{DBLP:conf/approx/RashtchianWZ20,SunWYZ19,ShiW019,BalcanLW019} have used linear algebraic queries like \emph{vector-matrix-vector query} and \emph{matrix-vector} queries. The \iporacle{} oracle is also motivated by these new query oracles. Notice the huge difference in power between \emph{matrix entry} query and \emph{vector-matrix-vector query} and \emph{matrix-vector} queries. Note that \iporacle{} is strictly weaker than these matrix queries but stronger than the \emph{matrix entry} query. We feel there is a need to study linear algebraic queries with intermediate power -- the \iporacle{} query fits in that slot.

%% file: lowerbounds-induced-subgraphs.tex
\section{A query model for induced subgraph problems}
\label{sec:induced}
\noindent
To the best of our knowledge, our work is a first attempt towards solving estimation problems in induced subgraphs. We start by showing a separation between {\sc local} query and {\sc induced degree} query using the problems of {\sc Edge Estimation} and {\sc Edge Sampling} in induced subgraph. \remove{We follow it up by showing, in Section~\ref{sec:induced_upper}, that {\sc induced degree} query on the whole unknown graph can simulate any {\sc local} query in any induced subgraph.} We now define \indedge and \indsamp. 

\defproblem{\indedge}{A parameter $\eps \in (0,1)$ and a subset $S$ of the vertex set $V$ of a graph $G$.}{A $(1 \pm \eps)$-approximation to the number of edges $E_{S}$ in the induced subgraph.}

\defproblem{\indsamp}{A parameter $\eps \in (0,1)$ and a subset $S$ of the vertex set $V$ of a graph $G$.}{Sample each edge $e \in E_S$ with probability between $\frac{1-\eps}{\size{E_S}}$ and $\frac{1+\eps}{\size{E_S}}$.}

One of the main contributions of this paper is to show, using a lower bound argument, that {\sc local} queries together with {\sc random edge} query are \emph{inefficient} for both \indedge and \indsamp. 
The lower bound results follow.
\begin{theo}[Lower bound for \indedge using {\sc local} queries]
\label{theo-lower-bound-induced--RESTATE}
    Let us assume that $s, m_s \in \N$ be such that $1\leq m_s\leq {s \choose 2}$ and the query algorithms have access to {\sc degree}, {\sc neighbor}, {\sc adjacency} and {\sc random edge} queries to an unknown graph $G = (V(G), E(G))$. Any query algorithm that can decide for all $S \subseteq V(G)$, with $\size{S} = \Theta(s)$, whether $\size{E_{S}} = m_s$ or $\size{E_{S}} = 2m_s$, with probability at least $2/3$, requires $\Omega\left( \frac{s^{2}}{m_s}\right)$  queries.
\end{theo}
\begin{theo}[Lower bound for \indsamp using {\sc local queries}]
\label{theo-lower-bound-induced-sampling-local-resate}
Let us assume that $s,m_s \in \N$ be such that $1\leq m_s\leq {s \choose 2}$ and the query algorithms have access to {\sc degree}, {\sc neighbor}, {\sc adjacency} and {\sc random edge} queries to an unknown graph $G = (V(G), E(G))$. Any query algorithm that for any $S \subseteq V(G)$, with $\size{S} = \Theta(s)$, samples the edges in $E_S$ $\epsilon$-almost uniformly~\footnote{Each edge in $E_S$ is sampled with probability between $(1-\eps)\frac{1}{\size{E_S}}$ and $(1+\eps)\frac{1}{\size{E_S}}$.}, with probability at least $99/100$, will require $\Omega\left( \frac{s^{2}}{m_{s}}\right)$  queries~\footnote{Let $U$ denote the uniform distribution on $E_S$. The lower bound even holds
even if the goal is to get a distribution that is $\epsilon$ close to $U$ with respect to $\ell_{1}$ distance.}. Note that $\epsilon \in (0,1)$ is given as an input to the algorithm. 
\end{theo}
\begin{rem}
\label{rem:separate}
When $S=V$, \indedge and \indsamp are {\sc Edge Estimation} and {\sc Edge Sampling} problems, respectively. Both {\sc Edge Estimation} and {\sc Edge Sampling} can be solved with high probability by using $\tTh \left({\size{V}^2}/{\size{E}}\right)$ {\sc adjacency} {queries~\cite{G2017}}. Notice that these bounds match the lower bounds. Contrast this with the fact that {\sc Edge Estimation} and {\sc Edge Sampling} can be solved with high probability by using
$\tTh\left({\size{V}}/{{\sqrt{\size{E}}}}\right)$ 
local queries, where each local query is either a {\sc degree} or a {\sc neighbor} or an {\sc adjacency} {query~\cite{GoldreichR08,EdenR18}}. Thus, we observe that for \indedge and \indsamp, the {\sc adjacency} query is as good as the entire gamut of {\sc local} queries and {\sc random edge} query. On a different note, our results on {\sc Bilinear Form Estimation} and {\sc Almost uniformly Sampling} using \iporacle{} query (see Table~\ref{table:graphresults1-main}) generalize the above mentioned results on {\sc Edge Estimation} and {\sc Edge Sampling} using local queries. {Note that \iporacle{} oracle is a natural query oracle for graphs where the unknown matrix is the adjacency matrix of a graph, and we will discuss that in Remark~\ref{rem:simu} that \iporacle{} query on the adjacency matrix graphs is stronger than the local queries.}
\end{rem}

In Section~\ref{sec:induce_lower}, we prove Theorems~\ref{theo-lower-bound-induced--RESTATE} and ~\ref{theo-lower-bound-induced-sampling-local-resate} by reduction from a problem in communication complexity. 
In Section~\ref{sec:induced_upper}, we discuss the way in which {\sc induced degree} query simulates local queries in any induced subgraph (see Remark~\ref{rem:simu}). This will imply that the lower bound results in Theorems~\ref{theo-lower-bound-induced--RESTATE} and ~\ref{theo-lower-bound-induced-sampling-local-resate} can be overcome if we have an access to {\sc induced degree} query to the whole graph (see Corollary~\ref{coro:induced_edge_ub}). However, the implication is more general and will be discussed in Section~\ref{sec:induced_upper}.

\remove{
\defproblem{\bipedge}{A parameter $\eps \in (0,1)$ and two disjoint subsets $A,B$ of the vertex set of a graph $G$.}{An $(1 \pm \eps)$-approximation to the number of edges in $E_{A,B}$.}\\

\defproblem{\bipsamp}{A parameter $\eps \in (0,1)$ and two disjoint subsets $A,B$ of the vertex set of a graph $G$.}{Sample each edge $e \in E_{A,B}$ with probability between $(1-\eps)\frac{1}{\size{E_{A,B}}}$ and $(1+\eps)\frac{1}{\size{E_{A,B}}}$.}
}



\subsection{Proofs of Theorems~\ref{theo-lower-bound-induced--RESTATE} and~\ref{theo-lower-bound-induced-sampling-local-resate}}\label{sec:induce_lower}
\label{sec:lbd_proofs}
The proofs of the lower bounds use communication complexity. We provide a rudimentary introduction to communication complexity in Appendix~\ref{sec:comm-comp}, and for more details see~\cite{KN97}. We will use the following problem in our lower bound proofs. 


\remove{\begin{defi}[{\sc Find-$k$-Intersection}]
\label{defi:find-intersection}
Let $k,N \in \N$ such that $k\leq N$. Let $S=\{({\bf x},{\bf y}) \in \{0,1\}^N \times \{0,1\}^N: \sum_{i=1}^N x_iy_i = k\}$. The {\sc Find-$k$-Intersection} function on $N$ bits is a partial function and is defined as $\FindIntersection_k^N: S \rightarrow \{0,1\}^N$, and is defined as $$\FindIntersection_k^N({\bf x},{\bf y}) = {\bf z}, \mbox{where}~ z_i=x_iy_i~ \mbox{for each}~ i \in [N].$$   
\end{defi}
Note that the objective is that at the end of the protocol Alice and Bob know ${\bf z}$.}

\begin{defi}[{\sc $k$-Intersection}]
\label{defi:k-intersection}
Let $k,N \in \N$ such that $k \leq N$. Let $S=\{({\bf x},{\bf y}): {\bf x},{\bf y}\in\{0,1\}^N, \sum_{i=1}^N x_iy_i=k~\mbox{or}~ 0\}$. The {\sc $k$-Intersection} function over $N$ bits is a partial function denoted by  $\mbox{\textsc{{\sc $k$-Intersection}}}:S   \rightarrow \{0,1\}$, and is defined as follows: $\textsc{{\sc $k$-Intersection}}({\bf x},{\bf y}) =1$ if $\sum_{i=1}^Nx_iy_i = k$~~\remove{Here $x_i$ and $y_i$ denote the $i$-the coordinate of ${\bf x}$ and ${\bf y}$.} and $0$, otherwise.
\remove{
\begin{align*}
    \textsc{{\sc $k$-Intersection}}({\bf x},{\bf y}) = 
    \begin{cases}
    1 & \tn{ if } \sum_{i=1}^Nx_iy_i = k\\
    0 & \tn{otherwise}
    \end{cases}
\end{align*}}
\end{defi}

\remove{\begin{theo}~\cite{KN97}
\label{theo:disj}
The randomized communication complexity of {\sc Disjointness} on $N$ bits is $\Omega(N)$.
\end{theo}}

\remove{Using standard reductions from {\sc Disjointness}, we can deduce the randomized communication complexities of {\sc $k$-Intersection} and {\sc Find-$k$-Intersection} as stated in Lemma~\ref{theo:kinter} and~\ref{theo:finding-intersection-hard}, respectively.}

\remove{\begin{lem}
\label{theo:finding-intersection-hard}
Let $k,N \in \N$ such that $k \leq cN$ for some constant $c<1$. The randomized communication complexity of {\sc Find-$k$-Intersection} function on $N$ bits is $\Omega(N)$.
\end{lem}}

\begin{lem}\cite{KN97}
\label{theo:kinter}
Let $k,N \in \N$ such that $k \leq N$. The randomized communication complexity of {\sc $k$-Intersection} function on $N$ bits is $\Omega\left(N/k\right)$. 
 \end{lem}

\begin{proof}[Proof of Theorem~\ref{theo-lower-bound-induced--RESTATE}]
We give a reduction from $m_s$-{\sc Intersection} problem over $N=s^2$ bits. Let $\boldvec{x} = (x_{ij}) \in \{0,1\}^N$ be such that $i,j \in [s]$. Similarly, let $\boldvec{y} \in \{0,1\}^N$. 
 It is promised that Alice and Bob will be given $\boldvec{x}$ and $\boldvec{y}$ such that there are either $0$ intersections or exactly $m_s$ intersections, i.e., either $\langle \boldvec{x}, \boldvec{y} \rangle = 0$ or $m_s$.
Now we define a graph $G_{(\boldvec{x},\boldvec{y})}(V(G),E(G))$ as follows where $\sqcup$ denotes disjoint union.
\begin{itemize}
    \item $\size{V(G)}=\Theta(s)$. $V(G)=S_A\sqcup S_B \sqcup T_A \sqcup T_B \sqcup C $ such that $S_A, \, S_B, \, T_A, \, T_B$ are independent sets and $\size{S_A}=\size{S_B}=\size{T_A}=\size{T_B}=s$ and  $\size{C}=\Theta(s)$. Note that $V(G)$ is independent of $\boldvec{x}$ and $\boldvec{y}$;
      \item The subgraph (of $G_{(\boldvec{x},\boldvec{y})}$) induced by $C$ is {a fixed} graph, independent of $\boldvec{x}$ and $\boldvec{y}$, having exactly $m_s$ edges. Also there are no edges in $G_{(\boldvec{x},\boldvec{y})}$ between the vertices of $C$ and $V(G)\setminus C$.
    \item The edges in the subgraph (of 
    $G_{(\boldvec{x},\boldvec{y})}$) induced by $V(G)\setminus 
    C=S_A \sqcup T_A \sqcup S_B \sqcup T_B$ depend on 
    $\boldvec{x}$ and $\boldvec{y}$ as follows. Let $S_A = \{s_i^A : i \in [s]\}$,  $T_A = \{t_i^A : i \in [s]\}$, $S_B = \{s_i^B : i \in [s]\}$ and $T_B = \{t_i^B : i \in [s]\}$. For  $i,j \in [s]$, if $x_{ij} = y_{ij} = 1$, then $(s_i^A, t_j^B) \in 
    E(G)$ and $(s_i^B, t_j^A) \in E(G)$. For  $i,j \in [s]$ if 
    either $x_{ij} = 0$ or $y_{ij} = 0$, then $(s_i^A, t_j^A) \in 
    E(G)$ and $(s_i^B, t_j^B) \in E(G)$;

\end{itemize}
The graph $G_{(\boldvec{x},\boldvec{y})}$ can be uniquely generated from $\boldvec{x}$ and $\boldvec{y}$. Moreover, \remove{no relevant information of $G_{(\boldvec{x},\boldvec{y})}$ can be derived by knowing exactly one of $\boldvec{x}$ or $\boldvec{y}$, that is,} Alice and Bob need to communicate to learn \emph{useful} information about $G_{(\boldvec{x},\boldvec{y})}$. \remove{
Let us have the following observation about graph $G_{(\boldvec{x},\boldvec{y})}$ that establishes that deciding whether the number of edges in the subgraph of $G_{(\boldvec{x},\boldvec{y})}$ induced by $S_A \cup T_B$ is $m_s$ or $2m_s$ is equivalent to deciding whether $\langle \boldvec{x}, \boldvec{y} \rangle = 0$ or $\langle \boldvec{x}, \boldvec{y} \rangle = m_s$, respectively (see particularly (iv) and (v) in the following observation).} Observation~\ref{obs:induce-edge-lb} follows from the construction that shows the relation between the number of edges in the subgraph induced by $S_A \sqcup T_B \sqcup C$ with $\langle \boldvec{x}, \boldvec{y} \rangle$, where $\boldvec{x},\boldvec{y} \in \{0,1\}^N$ are such that either $\langle \boldvec{x}, \boldvec{y} \rangle = 0$ or $\langle \boldvec{x}, \boldvec{y} \rangle = m_s$.

\begin{obs}\label{obs:induce-edge-lb}
    (i) $|S_A \sqcup T_B \sqcup C| = \Theta(s)$,
     (ii) irrespective of $x$ and $y$: $|E_{S_A}| = |E_{S_B}| = |E_{T_A}| = |E_{T_B}| = 0$, {also the degree of each vertex in $S_A\sqcup T_A \sqcup S_B \sqcup T_B$ is same (i.e., $s$)},
    (iii) if $\langle \boldvec{x}, \boldvec{y} \rangle = 0$, then $|E_{S_A \sqcup T_B \sqcup C}| = m_s$,
    (iv) if $\langle \boldvec{x}, \boldvec{y} \rangle = m_s$, then $|E_{S_A \sqcup T_B\sqcup C}|  = 2m_s$.
\end{obs}

The following observation completes the proof of the theorem.
\begin{obs}
Alice and Bob can deterministically determine answer for each local query to graph $G_{(\boldvec{x},\boldvec{y})}$ by communicating $\Oh(1)$ bits. \qedhere
\label{obs:queryimplement}
\end{obs}
\begin{proof}
\begin{description}
    \item[{\sc degree} query:] By Observation~\ref{obs:induce-edge-lb} (ii), the degree of every vertex in $V(G)\setminus C$ is $s$. \remove{irrespective of the inputs of Alice and Bob.} Also, the subgraph induced by $C$ is a fixed graph disconnected from the rest. That is Alice and Bob know the degree of every vertex in $C$. Therefore,  any {\sc degree} query can be simulated without any communication.
    
    \item[{\sc neighbor} query:] Observe that Alice and Bob can get the answer to any neighbor query involving a vertex in $C$ without any communication. Now,  consider the set $S_A$. The labels of the $j$ many neighbors of any vertex in  $s_i^A \in S_A$ are as follows: for $j \in [s]$, the $j$-th neighbor of $s_i^A$ is either $t_j^B$ or $s_j^A$ depending on whether $x_{ij}=y_{ij}=1$ or not, respectively. So, any {\sc neighbor} query involving vertex in $S_A$ can be answered by $2$ bits of communication. Similar arguments also hold for the vertices in $S_B \sqcup T_A \sqcup T_B$.
    \item[{\sc adjacency} query:] Observe that each adjacency query can be answered by at most $2$ bits of communication, and it can be argued like the {\sc neighbor} query.
    \item[{\sc random  edge} query:] By Observation~\ref{obs:induce-edge-lb} (ii), the degree of every vertex in $V(G)\setminus C$ is $s$ irrespective of the inputs of Alice and Bob. Also, they know the entire subgraph induced by the vertex set $C$. Also, $C$ is disconnected from the rest. Alice and Bob use shared randomness to sample a vertex in $V$ proportional to its degree. Let $r \in V$ be the sampled vertex. They again use shared randomness to sample an integer $j$ in $[d(v)]$ uniformly at random. Then they determine the $j$-th neighbor of $r$ using {\sc neighbor} query. Observe that this procedure simulates a {\sc random edge} query by using at most $2$ bits of communication.
\end{description}
\end{proof}
\end{proof}
\begin{proof}[Proof of Theorem~\ref{theo-lower-bound-induced-sampling-local-resate}]
For clarity, we prove the theorem for $\epsilon=1/4$. However, the proof can be extended for any $\epsilon \in (0,1/2)$.
We use the same set up and construction as in Theorem~\ref{theo-lower-bound-induced--RESTATE} with $S = S_A \sqcup S_B\sqcup C$.
Let $\cA$ be an algorithm that almost uniformly samples edges from the induced graph $G_S=(S, E_S)$ making $T$ queries, with probability $99/100$. 
Using $\cA$ we give another algorithm $\cA'$ that decides whether $\size{E_S} = m_s$ or $\size{E_S} = 2m_s$ by using $\Oh(T)$ queries, with probability at least $2/3$. From the reduction presented in Theorem~\ref{theo-lower-bound-induced--RESTATE}, Alice and Bob can use $\cA'$ to solve {\sc $m_s$-Intersection} over $N=s^2$ bits, and hence $T = \Omega(\frac{s^{2}}{m_s})$.

$\cA'$ runs $\cA$ $10$ times independently to obtain edges $e_1, \ldots, e_{10}$. Note that each edge $e_i$ is sampled almost uniformly. If at least one $e_i$ satisfies $e_i \in E_{S_A\sqcup T_B}$, then $\cA'$ reports that $\size{E_S} = 2m_s$. Otherwise, $\cA'$ reports that $\size{E_S} = m_s$. The query cost of $\cA'$ is $\Theta(T)$. 

If $\size{E_S} = m_s$, then there is no edge in the subgraph induced by $S_A\sqcup T_B$. So, in this case, all the edges reported by $\cA'$ are from the subgraph induced by $C$. Now consider when $\size{E_S} = 2m_s$. In this case, the subgraph induced by $S_A \sqcup T_B$ and $C$ have exactly $m_s$ edges each. So, by the assumption of the algorithm $\cA$, the probability that any particular $e_i$ is present in the subgraph induced by $S_A \sqcup T_B$ is at least $1/2-\epsilon=1/4$ (since, we are analyzing for $\epsilon =1/4$). So, under the conditional space that all the ten runs of $\cA$ succeed, the probability that none of the ten edges sampled by $\cA$ is from the subgraph induced by $S_A \sqcup T_B$ is at most $\left(1-1/4\right)^{10} < 1/10$. As each run of algorithm $\cA$ succeeds with probability at least $99/100$, all the ten runs of the algorithm $\cA$ succeeds with probability at least $9/10$. So,
the probability that algorithm $\cA'$ succeeds is at least $9/10\cdot (1-1/10)>2/3$.
%
\end{proof}

\subsection{A query model for induced subgraphs}
\label{sec:induced_upper}

We will first show that 
{\sc induced degree} query can simulate any {\sc local} query in any induced subgraph.

\begin{rem}\label{rem:simu}
Let us have an {\sc induced degree} query oracle access to an unknown graph $G(V,E)$. Consider any $X \subseteq V(G)$ and $G_X$, the subgraph of $G$ induced by $X$. Then
\begin{description}
\item[(i)] Any query to $G_X$, which is either a {\sc degree} or {\sc adjacency}, can be answered by one {\sc induced degree} query to $G$. 
\item[(ii)] Moreover, any {\sc neighbor} query to $G_X$ can be answered by $\Oh(\log |X|)$ many {\sc induced degree} query to $G$ by \emph{binary search}.
\end{description}
\end{rem}


The above remark together with the edge estimation result of Goldreich and Ron~\cite{GoldreichR08}, and edge sampling result of Eden and Rosenbaum~\cite{EdenR18}, gives us the following result as a corollary.
\begin{coro}[Upper bound for induced edge estimation and sampling]\label{coro:induced_edge_ub}
 Let us assume that the query algorithms
have access to {\sc induced degree} query to an unknown graph $G = (V (G), E(G))$. Then, there exists an algorithm that takes a subset $S \subseteq V(G)$ and $\eps \in (0,1)$ as inputs, and outputs a $(1 \pm \eps)$-approximation to $|E_ S|$, with high
probability, using $\tOh\left({|S|}/{\sqrt{E_S}}\right)$
 {\sc induced degree} queries to $G$. Also, there exists an algorithm that $\eps$-almost uniformly samples edges in $E_S$, with high probability, using $\tOh\left({|S|}/{\sqrt{E_S}}\right)$
 {\sc induced degree} queries to $G$.
\end{coro}

\begin{rem}
More generally, Remark~\ref{rem:simu} implies that any problem $\cP$ on a graph $G$ that can be solved by using $f\left(|V(G)|,|E(G)|\right)$ many local queries, can also be solved on any induced subgraph $G_{S}$, where $S \subseteq V(G)$, of $G$ by using $f\left(|V(G_S)|,|E(G_S)|\right) \cdot \Oh(\log |V(G_S)|)$ many {\sc induced degree} queries. 
\end{rem}

\remove{
\begin{coro}[{\sc Clique estimation} in induced subgraphs]
There exists an algorithm, that has {\sc induced degree} query access to an unknown graph $G$, takes $X\subseteq V(G) $ and a parameter $\eps \in (0,1)$ as input, reports an 
$(1\pm \eps)$-approximation to number of $h$-cliques~\footnote{$h$-cliques refers to cliques with $h$ many vertices} in $G_X$ with high probability, and makes $\widetilde{\Oh}\left(\frac{|X|}{(\# K_h)^{1/h}}+\min \left\{|E(G_X)|, \frac{|E(G_X)|^{t/2}}{\# K_h}\right\}\right)$~\footnote{$\# K_h$ denotes the number of $h$-cliques in $G_X$.} many {\sc induced degree} queries to $G$.
\end{coro}

Note that the above corollary directly follows from Lemma~\ref{}, and the clique estimation result due to Eden et al.~\cite{EdenRS18}.
}

%% file: bilinear.tex
\section{Bilinear form estimating and sampling entries of a matrix}
\label{sec:matrix}
\subsection{{Algorithm for Bilinear Form Estimation}}
\label{sec:algo-bilin}

To give the main ideas behind the algorithm for  {\sc Bilinear Form Estimation}, we will discuss, in this section, the algorithm for estimating $\mathbf{1}^{T}A\mathbf{1}$ using \iporacle{} access to $A$, with $A$ being symmetric. In Appendix~\ref{app:results} we show how the algorithm for this special case can be extended for the general problem of estimating $\mathbf{x}^T A \mathbf{y}$, where $A \in [[\rho]]^{n \times n}$, $\mathbf{x} \in [[\gamma_{1}]]^{n}$ and $\mathbf{y} \in [[\gamma_{2}]]^{n}$. We will give an outline of the proof of the following theorem. 

\begin{theo}\label{theo:algo_quad}
There exists a query algorithm for \onebfe that takes $\eps \in (0,1/2)$ as input and determines a $(1 \pm \eps)$-approximation to 
$\textbf{1}^TA \textbf{1}$ with high probability by using $\tOh\left(\frac{\sqrt{\rho} n}{\sqrt{\textbf{1}^TA \textbf{1}}}\right)$ many \iporacle{} queries to a symmetric matrix $A \in [[\rho]]^{n \times n}$.
Moreover, the algorithm only uses \iporacle{} query of the form $\dpd{A_{k*}}{{\bf u}}$ for some $k \in [n]$ and ${\bf u} \in \{0,1\}^n$.
\end{theo}
The algorithms for \onebilinear and \onealmostuni (Section~\ref{sec:sample}) will use a subroutine, which takes as input a given row $i \in [n]$ of $A$ and a non-empty set $S\subseteq [n]$, and outputs $A_{ij}$, where $j\in S$, with probability $A_{ij}/\left( \sum_{j \in S} A_{ij} \right)$.
\remove{
\begin{algorithm}[ht]
\caption{\regr($\bx,i$)}
\label{algo_cvd}
\KwIn{A vector $\bx \in \{0,1\}^n$ such that the $1$'s in $\bx$ are consecutive and the number of $1$'s is a power of $2$, an integer $i \in [n]$ and \iporacle{} access to a matrix $A$.}
\KwOut{Ordered pair $(i,j)$ with probability $\frac{{A_{ij}}\cdot{ x_j}}{\dpd{ A_{i,*}}{\bx}}$.}
\Begin
	{
	\If{(the number of $1$'s in $\bx$ is 1)}
	{Report the ordered pair $(i,j^{*})$ where $x_{j^*}=1$.}
	
	\Else
	{
	Form a vector ${\bf{y}}$ (${\bf{z}}$) in $\{0,1\}^n$ by setting second (first) half of the nonzero elements in ${\bf{y}}$ (${\bf{z}}$) to $0$ and keeping the remaining elements unchanged.\\ 
	
	Determine $\dpd{A_{i,*}}{\bf{y}}$ and $\dpd{A_{i,*}}{\bf{z}}$. \\
	
	With probability $\frac{\dpd{A_{i,*}}{ \bf{y}}}{\dpd{ A_{i,*}}{ \bf{x}}}$
	report \regr($\bf{y},i$) and with probability $\frac{\dpd{A_{i,*}}{ \bf{z}}}{\dpd{A_{i,*}}{ \bf{x} }}$
	report \regr($\bf{z},i$).
	}
}
\end{algorithm}}
\label{app:regr-code}
\begin{algorithm}[h]
\caption{\regr($\bx,i$)}
\label{algo_cvd}
\KwIn{A vector $\bx \in \{0,1\}^n$ such that the $1$'s in $\bx$ are consecutive and the number of $1$'s is a power of $2$, an integer $i \in [n]$ and \iporacle{} access to a matrix $A$.}
\KwOut{Ordered pair $(i,j)$ with probability $\frac{{A_{ij}}\cdot{ x_j}}{\dpd{ A_{i*}}{\bx}}$.}
\Begin
	{
	\If{(the number of $1$'s in $\bx$ is 1)}
	{Report the ordered pair $(ij^{*})$ where $x_{j^*}=1$.}
	
	\Else
	{
	Form a vector ${\bf{y}}$ (${\bf{z}}$) in $\{0,1\}^n$ by setting second (first) half of the nonzero elements in ${\bf{y}}$ (${\bf{z}}$) to $0$ and keeping the remaining elements unchanged.\\ 
	
	Determine $\dpd{A_{i*}}{\bf{y}}$ and $\dpd{A_{i*}}{\bf{z}}$. \\
	
	With probability $\frac{\dpd{A_{i*}}{ \bf{y}}}{\dpd{ A_{i*}}{ \bf{x}}}$
	report \regr($\bf{y},i$) and with probability $\frac{\dpd{A_{i*}}{ \bf{z}}}{\dpd{A_{i*}}{ \bf{x} }}$
	report \regr($\bf{z},i$).
	}
}
\end{algorithm}
\begin{obs}
\label{lem:random}
{There exists an algorithm \regr{} (See Algorithm~\ref{algo_cvd}) that takes $i \in [n]$ and $\mathbf{x} \in \{0,1\}^{n}$ as inputs, outputs $A_{ij}$ with probability ${A_{ij}x_{j}}/{\left( \sum_{j \in [n]} A_{ij}x_j \right)}$ by using $\Oh(\log n)$ many \iporacle{} queries to matrix $A$.}
\end{obs}
We will now discuss in the following paragraphs the details of the algorithm (Algorithm~\ref{algo_onebfe}) for estimating $\mathbf{1}^{T} A\mathbf{1}$. The ingredients,  to prove the correctness of Algorithm~\ref{algo_onebfe}, is formally stated  in Lemma~\ref{onebfe_technical}. The approximation guarantee of Algorithm~\ref{algo_onebfe}, which matches the guarantee mentioned in Theorem~\ref{theo:algo_quad}, is given in Claim~\ref{cl:approx}.
\paragraph*{Partition of rows of $\mathbf{A}$ induced by $\epsilon$}
 Given $\epsilon$ as input, we argue that the rows of the symmetric matrix ${ A}$ can be partitioned into ``buckets'' such that the total number of buckets is small and every row in a particular bucket has approximately the same total weight. Consider a partition of $[n]$, that corresponds to the set of the indices of the rows of the symmetric matrix 
${A}$, into buckets with the property that all $j$s present in a particular bucket $B_i$ have approximately the same value of $\dpd {A_{j*}} {\textbf{1}}$. 
Let 
$t = \lceil \log_{1+\beta} (\rho n) \rceil + 1$, where $\beta \leq \eps/8$. 
For $i \in [t]$, we define the set $B_i := \left\{j \in [n]~:~(1+\beta)^{i-1} \leq \dpd {A_{j*}} {\textbf{1}} < (1+\beta)^i\right\}$. Since 
$A_{ij} \leq \rho$, the maximum number of such buckets $B_i$ required are at most
$t = \lceil \log_{1+\beta} (\rho n) \rceil + 1$.
Now consider the following fact that will be used in our analysis.
\begin{fact}
For every $i \in [t]$, $(1+\beta)^{i-1} \size{B_i} \leq \sum_{j \in B_i} \dpd {A_{j*}} {\textbf{1}} < (1+\beta)^i \size{B_i}$.
\label{fact:bucket}
\end{fact}

Based on the number of rows in a bucket, we classify the buckets to be either {\em large} or {\em small}. To define the large and small buckets, we require a lower bound $\ell$ on the value of $m=\oneaone$. Moreover, let us assume that, $m/6\leq \ell \leq m$. However, this restriction can be removed by using standard techniques from property testing. For details, see Appendix~\ref{sec:app-matrix}.

 \begin{defi}
 \label{defi:large-small}
We fix a threshold 
$\theta = \frac{1}{t} \cdot \frac{1}{n} \sqrt{\frac{\epsilon}{8} \cdot \frac{\ell}{\rho}}$. For $i \in [t]$, we define the set $B_i$ to be a \emph{large bucket} if  
$|B_i| \geq \theta n$.
Otherwise, the set $B_i$ is defined to be a \emph{small bucket}. Thus, the set of large buckets $L$ is defined as 
$L = \{i \in [t]:~|B_i| \geq \theta n \}$, 
and $[t] \setminus L$ is the set of small buckets.
\end{defi}

Let $V,U \subseteq [n]$ be the sets of indices of rows that lie in large and small buckets, respectively. For $I \subseteq [n]$, let ${\bf{x}}_I$ denote the sub-vector of $\bf{x}$ induced by the indices present in $I$. Similarly, for $I,J \subseteq [n]$, let $A_{IJ}$ denote the sub-matrix of $A$ where the rows and columns are induced by the indices present in $I$ and $J$, respectively. 
Observe that,
$\oneaone = {{\bf{1}}_{V}}^T A_{VV}{{\bf{1}}_V}+{{\bf{1}}_{V}}^T A_{VU}{{\bf{1}}_U} + {{\bf{1}}_{U}}^T A_{UV}{{\bf{1}}_V}+ {{\bf{1}}_{U}}^T A_{UU}{{\bf{1}}_U}$.
\paragraph*{$2$-Approximation of $\mathbf{1^T A 1}$} 
Note that at this point we know $\beta$ and, upon querying $\dpd{A_{j*}}{\vecone}$, we can determine the bucket to which $j$ belongs, for $j \in [n]$. The algorithm begins {by sampling a subset $S$ of} rows of $A$, such that $|S| = K$, independently and uniformly at random with replacement, and for each sampled row $j$, the
algorithm determines $\dpd{A_{j*}}{\vecone}$ by using \iporacle{} oracle. This determines the bucket in which each sampled row belongs. Depending on the number of sampled rows present in different buckets, our algorithm classifies each bucket as either large or small. Let $\Tilde{V}$ and $\Tilde{U}$ be the indices of the rows present in large and small buckets, respectively. Note that the algorithm does not find $\Tilde{V}$ and $\Tilde{U}$ explicitly -- these are used only for analysis purpose.  

Observe that,
$\oneaone = {{\bf{1}}_{\tV}}^T A_{\tV \tV}{{\bf{1}}_{\tV}}+{{\bf{1}}_{\tV}}^T A_{\tV \tU}{{\bf{1}}_{\tU}} + {{\bf{1}}_{\tU}}^T A_{\tU \tV}{{\bf{1}}_{\tV}}+ {{\bf{1}}_{\tU}}^T A_{\tU \tU}{{\bf{1}}_{\tU}}$.  
We can show that ${{\bf{1}}_{\tU}}^T A_{\tU \tU}{{\bf{1}}_\tU}$ is at most $\frac{\eps}{4} \ell$, where $\ell$ is a lower bound on $\oneaone$. Thus,
$\oneaone \approx {{\bf{1}}_{\tV}}^T A_{\tV \tV}{{\bf{1}}_{\tV}}+{{\bf{1}}_{\tV}}^T A_{\tV \tU}{{\bf{1}}_{\tU}} + {{\bf{1}}_{\tU}}^T A_{\tU \tV}{{\bf{1}}_{\tV}}$.

Lemma~\ref{onebfe_technical} shows that for a sufficiently large $K$, with high probability, the fraction of rows in any large bucket is approximately preserved in the sampled set of rows. Also observe that we know tight (upper and lower) bounds on $\dpd {A_{j*}} {\textbf{1}}$ for every row $j$, where $j \in \tV$. Thus, {the random sample of $S$ rows}, such that $|S| = K$, approximately preserves ${{\bf{1}}_{\tV}}^T A_{\tV \tV}{{\bf{1}}_{\tV}}+{{\bf{1}}_{\tV}}^T A_{\tV \tU}{{\bf{1}}_{\tU}}$. Observe that this is already $2$-approximation of $\mathbf{1^TA1}$.
\paragraph*{Using \regr{} for tight approximation} 
In order to get a $(1 \pm \eps)$-approximation to $\oneaone$, we need to 
estimate ${{\bf{1}}_{\tU}}^T A_{\tU \tV}{{\bf{1}}_\tV}$, which is same as estimating ${{\bf{1}}_{\tV}}^T A_{\tV 
 \tU}{{\bf{1}}_\tU}$ since $A$ is a symmetric matrix. We estimate ${{\bf{1}}_{\tV}}^T A_{\tV 
\tU}{{\bf{1}}_\tU}$, that is, the sum of $A_{ij}$s such that $i \in \Tilde{V}$ and $j \in \Tilde{U}$, as follows. For each bucket $B_i$ that is declared as large by the algorithm, we select
\emph{enough} number of rows randomly with replacement from {$S_i=S \cap B_i$}, invoke \regr{} for each selected row in $S_i$ and increase the count by $1$ if the element $A_{ij}$ reported by \regr{} be such that $j \in \Tilde{U}$. A formal description of our algorithm is given in Algorithm~\ref{algo_onebfe}. Now, we focus on the correctness proof of our algorithm for \onebfe.

\begin{algorithm}

\caption{\onebfe($\ell$, $\epsilon$)}
\label{algo_onebfe}
\KwIn{An estimate $\ell$ for $\vecone^T A \vecone$ and $\epsilon \in \left(0,1/2\right)$.} 
\KwOut{$\widehat{m}$, which is a $(1 \pm \eps)$-approximation of
$\textbf{1}^TA \textbf{1}$.}
\Begin
	{
	Independently select $K = \Theta\left(\frac{\sqrt{\rho}n}{\sqrt{\ell}} \cdot \epsilon^{-4.5} \cdot \log^2(\rho n) \cdot \log (1/\epsilon) \right)$ rows of $A$ uniformly at random and let $S$ denote the multiset of the selected indices (of rows) sampled. For $i \in [t]$, let $S_i = B_i \cap S$. \\
	Let $\Tilde{L} = \Big\{i : \frac{|S_i|}{|S|} \geq \frac{1}{t} \cdot \frac{1}{n} \sqrt{\frac{\epsilon}{6} \cdot \frac{\ell}{\rho}}  \Big\}$. Note that $\Tilde{L}$ is the set of buckets that the algorithm declares to be large. Similarly, $ [t] \setminus \Tilde{L}$ is the set of buckets declared to be small by the algorithm. \\
	For every $i \in \Tilde{L}$, select $|S_i|$ samples uniformly at random from $S_i$, with replacement, and let $Z_i$  be the set of samples obtained. For each $z \in Z_i$, make a \regr($z,\vecone$) query and let $A_{zk_z}=~$\regr($z,\vecone$). Let $Y_z$ be a random variable that takes value $1$ if $k_z \in \Tilde{U}$ and $0$, otherwise.
	
	// If we take $Z_i=S_i$, then also the correctness can be proved. But comparatively, the correctness proof is slightly clean as because of the way we are generating $Z_i$.
	
   Determine $\Tilde{\alpha_i} = \frac{\sum_{{z \in Z_i}} Y_z}{|S_i|}$.

	Output
   $
   	\widehat{m} = \frac{n}{K} \sum_{i \in \Tilde{L}} (1 + \Tilde{\alpha}_i) \cdot |S_i| \cdot (1 + \beta)^i.
 $
}
\end{algorithm}

To prove that $\hat{m}$ is a $(1\pm \eps)$-approximation of $m=\oneaone$, we need the following definition and the technical Lemma~\ref{onebfe_technical}.
\begin{defi}
\label{defi:alpha-i}
For $i \in L$, $\alpha_i $ is defined as $\frac{\sum_{u \in B_i} \dpd {A_{u*}} {\textbf{1}_{\Tilde{U}}}}{\sum_{u \in B_i} \dpd {A_{u*}} {\textbf{1}}}$.
\end{defi}
\begin{lem}
\label{onebfe_technical}
For a suitable choice of constant in $\Theta(\cdot)$ for selecting $K$ samples in Algorithm~\ref{algo_onebfe}, the followings hold with high probability:
\remove{
(i) For each $i \in L$, $  (1 - \frac{\epsilon}{4})\frac{|B_i|}{n} \leq \frac{|S_i|}{K} \leq (1 + \frac{\epsilon}{4}) \frac{|B_i|}{n}$; (ii) For each $i \in [t] \setminus L$, $
         \frac{|S_i|}{K} < \frac{1}{t} \cdot \frac{1}{n} \sqrt{ \frac{\epsilon}{6}\cdot \frac{\ell}{\rho} }$; (iii) $ |\Tilde{U}| < \sqrt{ \frac{\epsilon}{4} \cdot \frac{\ell}{\rho}} $, where $\Tilde{U} = \{j \in B_i : i \in [t] \setminus \Tilde{L} \}$;
         (iv) For every $i \in \Tilde{L}$, 
    {\bf (a)} if $\alpha_i \geq \frac{\epsilon}{8}$, then $(1 - \frac{\epsilon}{4}) \alpha_i \leq \Tilde{\alpha_i} \leq (1 + \frac{\epsilon}{4}) \alpha_i$, and {\bf (b)} if $\alpha_i < \epsilon/8$, then $\Tilde{\alpha}_i < \epsilon/4$.}
    {
\begin{description}
\item[(i)]
    For each $i \in L$, $  (1 - \frac{\epsilon}{4})\frac{|B_i|}{n} \leq \frac{|S_i|}{K} \leq (1 + \frac{\epsilon}{4}) \frac{|B_i|}{n}$.
    
\item[(ii)]
    For each $i \in [t] \setminus L$, 
    $\frac{|S_i|}{K} < \frac{1}{t} \cdot \frac{1}{n} \sqrt{ \frac{\epsilon}{6}\cdot \frac{\ell}{\rho} }$.
    
\item[(iii)]
    We have $ |\Tilde{U}| < \sqrt{ \frac{\epsilon}{4} \cdot \frac{\ell}{\rho}} $, where $\Tilde{U} = \{j \in B_i : i \in [t] \setminus \Tilde{L} \}$.

\item[(iv)]
    For every $i \in \Tilde{L}$, 
    {\bf (a)} if $\alpha_i \geq \frac{\epsilon}{8}$, then $(1 - \frac{\epsilon}{4}) \alpha_i \leq \Tilde{\alpha_i} \leq (1 + \frac{\epsilon}{4}) \alpha_i$, and {\bf (b)} if $\alpha_i < \epsilon/8$, then $\Tilde{\alpha}_i < \epsilon/4$.
\end{description}}
\end{lem}
 
\begin{proof}
\begin{itemize}
\item[(i)] Recall that $K=|S|= \Theta\left(\frac{\sqrt{\rho}n}{\sqrt{\ell}} \cdot \epsilon^{-4.5} \cdot \log^2(\rho n) \cdot \log (1/\epsilon) \right)$. Observe that $\E[|S_i|]=\frac{|S|}{n}|B_i|$. Here $i\in L$. By the definition of $L$ (See Definition~\ref{defi:large-small}), 
$|B_i| \geq \frac{1}{t} \cdot  \sqrt{\frac{\epsilon}{8} \cdot \frac{\ell}{\rho}}$. So,  $\E[|S_i|]\geq \frac{|S|}{nt}\sqrt{\frac{\epsilon}{8} \cdot \frac{\ell}{\rho}}$. Using the facts that $|S|= \Theta\left(\frac{\sqrt{\rho}n}{\sqrt{\ell}} \cdot \epsilon^{-4.5} \cdot \log^2(\rho n) \cdot \log (1/\epsilon) \right)$ and $t=\lceil \log _{1+\eps/8} (\rho n) \rceil$, then applying Chernoff bound as mentioned in Lemma~\ref{lem:chernoff}(ii) in Appendix~\ref{sec:prob}, we get the desired result.
\item[(ii)] In this case, $\E[|S_i|]< \frac{|S|}{nt}\sqrt{\frac{\epsilon}{8} \cdot \frac{\ell}{\rho}}$. Now applying Chernoff bound as mentioned in part (a) of Lemma~\ref{lem:chernoff}(ii) in Appendix~\ref{sec:prob}, we get the the desired result.

{
\item[(iii)] By the definition of $\Tilde{U}$, $
\size{   \Tilde{ U}} = \Big| \left\{j \in B_i : \frac{|S_i|}{|S|} <  \frac{1}{t} \cdot \frac{1}{n} \sqrt{\frac{\epsilon}{6} \cdot \frac{\ell}{\rho}} \right\} \Big|$. Applying Lemma~\ref{onebfe_technical}(i) and the definition of $L$, we get $\size{\Tilde{U}}       \leq \Big| \left\{j \in B_i : \frac{|B_i|}{n} < (1 - \frac{\epsilon}{4})^{-1}  \cdot \frac{1}{t} \cdot \frac{1}{n} \sqrt{\frac{\epsilon}{6} \cdot \frac{\ell}{\rho}} \right\} \Big|$. 
As there are at most $t$ many buckets,
$$
    \size{\Tilde{U}}   < \Big| \left\{j \in B_i : |B_i| < \frac{1}{t}  \sqrt{\frac{\epsilon}{4} \cdot \frac{\ell}{\rho}} \right\} \Big|  \leq
        \sqrt{ \frac{\epsilon}{4} \cdot \frac{\ell}{\rho}} .
$$
}

\item[(iv)] From the description of the algorithm, for every $i \in \Tilde{L}$, we select $|S_i|$ many samples uniformly at random from $S_i$, with replacement, and let $Z_i$  be the set of samples obtained. For each $z \in Z_i$, we make a \regr($z,\vecone$) query and let $A_{zk_z}=$ \regr($z,\vecone$). Let $Y_z$ be a random variable that takes value $1$ if $k_z \in \Tilde{U}$ and $0$, otherwise. Also, $\Tilde{\alpha_i} = \frac{\sum_{{z \in Z_i}} Y_z}{|S_i|}$.

Using the fact that we choose $S$ independently and uniformly at random, set $S_i=S \cap B_i$ and sample the elements in $Z_i$ from $S_i$, we get
$$
    \E[Y_z]=\frac{\size{S_i}}{\size{B_i}}\cdot\frac{1}{\size{S_i}}\sum_{u \in B_i}\frac{\dpd{A_{u*}}{{\vecone}_{\tilde{U}}}}{\dpd{A_{u*}}{{\vecone}}}=\frac{1}{\size{B_i}}\cdot\sum_{u \in B_i}\frac{\dpd{A_{u*}}{{\vecone}_{\tilde{U}}}}{\dpd{A_{u*}}{\vecone}}.
$$

So, $\E[\tilde{\alpha_i}]=\sum_{u \in B_i} \frac{1}{\size{B_i}}\cdot\frac{\dpd{A_{u*}}{{\vecone}_{\tilde{U}}}}{\dpd{A_{u*}}{\vecone}}$.
Since $u \in B_i$, $(1+\beta)^{i-1} \leq \dpd {A_{u*}}{\textbf{1}} < (1+\beta)^i$. Also, by Fact 1,
$$
    \size{B_i}(1+\beta)^{i-1} \leq\sum_{u \in B_i}\dpd{A_{u*}}{{\vecone}} \leq \size{B_i}(1+\beta)^{i}.
$$
Thus, 
$$ 
    \frac{
\sum_{u \in B_i} \dpd {A_{u*}} {\textbf{1}_{\Tilde{U}}}
    	 }
    	 {|B_i| (1+\beta)^i} 
    \leq \mathbb{E} [ \Tilde{\alpha_i}] \leq      	
    \frac{ 
    	 \sum_{u\in B_i} \dpd{A_{u*}} {\textbf{1}_{\tilde{U}}}    	 }
    	 {|B_i| (1+\beta)^{i-1} },
$$ 
or
\begin{eqnarray*}
    	\frac{1}{1+\beta}\cdot  \frac{
\sum_{u \in B_i} \dpd {A_{u*}} {\textbf{1}_{\Tilde{U}}}
    	 }
    	 { \dpd{A_{u*}}{{\vecone}}} 
    &\leq& \mathbb{E} [ \Tilde{\alpha_i}] \leq      	
    \frac{ 
    	 \sum_{u\in B_i} \dpd{A_{u*}} {\textbf{1}_{\tilde{U}}}    	 }
    	 {\dpd{A_{u*}}{\vecone} }(1+\beta)
    	  \end{eqnarray*}
    	 Using $\alpha_i=\frac{\sum_{u \in B_i} \dpd {A_{u*}} {\textbf{1}_{\Tilde{U}}}}{\sum_{u \in B_i} \dpd {A_{u*}} {\textbf{1}}}$ (Definition~\ref{defi:alpha-i}) and the fact that $\beta \leq \eps/8$, we get
    	 $$
            \left(1-\frac{\eps}{4}\right) \alpha_i \leq \mathbb{E} [ \Tilde{\alpha_i}] \leq \left(1+\frac{\eps}{4} \right) \alpha_{i}.
        $$
Now the proof follows from Chernoff bound (Lemma~\ref{lem:chernoff} (ii), Appendix~\ref{sec:prob}).
\end{itemize}
\end{proof}

 Now, we have all the ingredients to show the following claim, which shows that $\hat{m}$ is a $(1 \pm \eps)$-approximation of $m=\oneaone.$

\begin{cl}
\label{cl:approx}
With high probability, we have, 
\begin{itemize}
\item[(i)] $ \widehat{m} \geq \left(1-\frac{\eps}{2}\right)\left(m-\frac{\eps}{4}\ell\right)$,  and 
\item[(ii)] $\widehat{m}\leq  \left(1+\frac{3\eps}{4}\right)m$, where $m={\bf 1}^TA{\bf 1}$.
\end{itemize}
\end{cl}
\begin{proof}
 \begin{itemize}
 \item[(i)]

Recall that $\widehat{m} = \frac{n}{K} \sum_{i \in \Tilde{L}} (1 + \Tilde{\alpha}_i) \cdot |S_i| \cdot (1 + \beta)^i$ is the estimate returned by Algorithm~\ref{algo_onebfe}. Using Lemma~\ref{onebfe_technical} (i), we have 
\begin{eqnarray*}
    \widehat{m} &\geq&  \sum_{i \in \Tilde{L}} (1 + \Tilde{\alpha}_i) 
                \left(1 - \frac{\epsilon}{4}\right)  (1+\beta)^i |B_i| \\
                &=&  (1 - \frac{\epsilon}{4}) \sum_{i \in \Tilde{L}} (1 + \Tilde{\alpha}_i) (1+\beta)^i |B_i| \\
                &=& \left(1 - \frac{\epsilon}{4}\right) \Big( \sum_{\substack{i \in \Tilde{L} \\ \alpha_i \geq \epsilon/8}} (1 + \Tilde{\alpha}_i) (1+\beta)^i |B_i| + \sum_{\substack{i \in \Tilde{L} \\ \alpha_i < \epsilon/8}} (1 + \Tilde{\alpha}_i) (1+\beta)^i |B_i| \Big) \\
                &\geq&  \left(1 - \frac{\epsilon}{4}\right) \Big( \sum_{\substack{i \in \Tilde{L} \\ \alpha_i \geq \epsilon/8}} \big(1 + (1 - \frac{\epsilon}{4})\alpha_i \big) (1+\beta)^i |B_i| + \sum_{\substack{i \in \Tilde{L} \\ \alpha_i < \epsilon/8}} (1 - \frac{\epsilon}{4}) (1 + \frac{\epsilon}{4}) (1+\beta)^i |B_i| \Big) \\
                &>&  \left(1 - \frac{\epsilon}{4}\right) \Big( \sum_{\substack{i \in \Tilde{L} \\ \alpha_i \geq \epsilon/8}} \big((1 - \frac{\epsilon}{4}) (1 + \alpha_i) \big) (1+\beta)^i |B_i| + \sum_{\substack{i \in \Tilde{L} \\ \alpha_i < \epsilon/8}} (1 - \frac{\epsilon}{4}) (1 + \alpha_i) (1+\beta)^i |B_i| \Big) \\
                &=&  \left(1 - \frac{\epsilon}{4}\right)^2  \sum_{i \in \Tilde{L}} (1 + \alpha_i)  (1+\beta)^i |B_i| \\
                &\geq&  \left(1 - \frac{\epsilon}{4}\right)^2  \sum_{i \in \Tilde{L}} \Big( (1 + \alpha_i)  \sum_{k \in B_i} \dpd{A_{k*}} {\textbf{1}}  \Big)
\end{eqnarray*}

Using $ \sum_{k \in B_i} \dpd{A_{k*}} {\textbf{1}_{  \Tilde{T}}} = \alpha_i \sum_{k \in B_i} \dpd{A_{k*}} {\textbf{1}} $ we have:
\begin{eqnarray*}
     \widehat{m} &\geq&  \left(1 - \frac{\epsilon}{4}\right)^2  
                 \sum_{i \in \Tilde{L}} \Big( (1 + \alpha_i)  \sum_{k \in B_i} \dpd{A_{k*}} {\textbf{1}} \Big) \\
                 &=&  \left(1 - \frac{\epsilon}{4}\right)^2
                 \sum_{i \in \Tilde{L}} \Big( \sum_{k \in B_i} \dpd{A_{k*}} {\textbf{1}}  + \alpha_i \sum_{k \in B_i} \dpd{A_{k*}} {\textbf{1}}  \Big) \\
                 &=&  \left(1 - \frac{\epsilon}{4}\right)^2
                 \sum_{i \in \Tilde{L}} \Big( \sum_{k \in B_i} \dpd{A_{k*}} {\textbf{1}_{\Tilde{L}}} + \sum_{k \in B_i} \dpd{A_{k*}} {\textbf{1}_{\Tilde{T}}} + \alpha_i \sum_{k \in B_i} \dpd{A_{k*}} {\textbf{1}} \Big) \\
                 &=& \left(1 - \frac{\epsilon}{4}\right)^2
                 \sum_{i \in \Tilde{L}} \Big( \sum_{k \in B_i} \dpd{A_{k*}} {\textbf{1}_{\Tilde{L}}} + 2 \sum_{k \in B_i} \dpd{A_{k*}} {\textbf{1}_{\Tilde{T}}} \Big) \\
                 &=& \left(1 - \frac{\epsilon}{4}\right)^2
                 \Big( \sum_{i \in \Tilde{L}}  \sum_{k \in B_i} \dpd{A_{k*}} {\textbf{1}_{\Tilde{L}}} + 2 \sum_{i \in \Tilde{L}} \sum_{k \in B_i} \dpd{A_{k*}} {\textbf{1}_{\Tilde{T}}} \Big) \\
                 &=& \left(1 - \frac{\epsilon}{4}\right)^2
                 \Big(  \textbf{1}_{\Tilde{V}}^T A_{\Tilde{V}, \Tilde{V}} \textbf{1}_{\Tilde{V}} + 2 \textbf{1}_{\Tilde{V}}^T A_{\Tilde{V}, \Tilde{U}} \textbf{1}_{\Tilde{U}} \Big)\\
                 &=&  \left(1 - \frac{\epsilon}{4}\right)^2
                 \Big(  \textbf{1}^T A \textbf{1} - \textbf{1}_{\Tilde{U}}^T A_{\Tilde{U}, \Tilde{U}} \textbf{1}_{\Tilde{U}} \Big)
\end{eqnarray*}
Since $|\Tilde{U}| < \sqrt{ \frac{\epsilon}{4} \cdot \frac{\ell}{\rho}}$ (From Lemma~\ref{onebfe_technical} (iii)) and $\forall$ $i,j \in [n]$, $|A_{ij}| \leq \rho$ we have
  $  \textbf{1}_{\Tilde{U}}^T A_{\Tilde{U}, \Tilde{U}} \textbf{1}_{\Tilde{U}} \leq \frac{\epsilon}{4} \cdot \ell $. 
So, we have $\widehat{m} \geq \left(1 - \frac{\epsilon}{4}\right)^2  \left(m-\frac{\eps}{4}\ell\right)$.

\item[(ii)]Using Lemma~\ref{onebfe_technical} (i), we have 
\begin{eqnarray*}
    \widehat{m} &\leq&  \sum_{i \in \Tilde{L}} (1 + \Tilde{\alpha}_i) (1 + \frac{\epsilon}{4})  (1+\beta)^i |B_i| \\
                &=&  (1 + \frac{\epsilon}{4}) \sum_{i \in \Tilde{L}} (1 + \Tilde{\alpha}_i) (1+\beta)^i |B_i| \\
                &=&  (1 + \frac{\epsilon}{4}) \Big( \sum_{\substack{i \in \Tilde{L} \\ \alpha_i \geq \epsilon/8}} (1 + \Tilde{\alpha}_i) (1+\beta)^i |B_i| + \sum_{\substack{i \in \Tilde{L} \\ \alpha_i < \epsilon/8}} (1 + \Tilde{\alpha}_i) (1+\beta)^i |B_i| \Big) \\
                &\leq&  (1 + \frac{\epsilon}{4}) \Big( \sum_{\substack{i \in \Tilde{L} \\ \alpha_i \geq \epsilon/8}} \big(1 + (1 + \frac{\epsilon}{4})\alpha_i \big) (1+\beta)^i |B_i| + \sum_{\substack{i \in \Tilde{L} \\ \alpha_i < \epsilon/8}} (1 + \frac{\epsilon}{4}) (1+\beta)^i |B_i| \Big) \\
                &<&  (1 + \frac{\epsilon}{4}) \Big( \sum_{\substack{i \in \Tilde{L} \\ \alpha_i \geq \epsilon/8}} \big((1 + \frac{\epsilon}{4}) (1 + \alpha_i) \big) (1+\beta)^i |B_i| + \sum_{\substack{i \in \Tilde{L} \\ \alpha_i < \epsilon/8}} (1 + \frac{\epsilon}{4}) (1 + \alpha_i) (1+\beta)^i |B_i| \Big) \\
                &=&  (1 + \frac{\epsilon}{4})^2  \sum_{i \in \Tilde{L}} (1 + \alpha_i)  (1+\beta)^i |B_i| \\
                &\leq&  (1 + \frac{\epsilon}{4})^2  (1+\beta) \sum_{i \in \Tilde{L}} \Big( (1 + \alpha_i)  \sum_{k \in B_i} \dpd{A_{k*}} {\textbf{1}} \Big)
\end{eqnarray*}

Since $\beta \leq \frac{\epsilon}{8}$ and $ \sum_{k \in B_i} \dpd{A_{k*}} {\textbf{1} _{  \Tilde{U}}} = \alpha_i \sum_{k \in B_i} \dpd{A_{k*}} {\textbf{1}} $ we have:
\begin{eqnarray*}
     \widehat{m} &\leq&  (1 + \frac{\epsilon}{4})^2  (1+\frac{\epsilon}{8}) 
     \sum_{i \in \Tilde{L}} \Big( (1 + \alpha_i)  \sum_{k \in B_i} \dpd{A_{k*}} {\textbf{1}}  \Big) \\
                 &\leq&  (1 + \frac{3\epsilon}{4})
                 \sum_{i \in \Tilde{L}} \Big( \sum_{k \in B_i} \dpd{A_{k*}} {\textbf{1}}  + \alpha_i \sum_{k \in B_i} \dpd{A_{k*}} {\textbf{1}}  \Big) \\
                 &=& (1 + \frac{3\epsilon}{4})
                 \sum_{i \in \Tilde{L}} \Big( \sum_{k \in B_i} \dpd{A_{k*}} {\textbf{1} _{\Tilde{V}}} + \sum_{k \in B_i} \dpd{A_{k*}} {\textbf{1}_{\Tilde{U}}} + \alpha_i \sum_{k \in B_i} \dpd{A_{k*}} {\textbf{1}}  \Big) \\
                 &=&  (1 + \frac{3\epsilon}{4})
                 \sum_{i \in \Tilde{L}} \Big( \sum_{k \in B_i} \dpd{A_{k*}} {\textbf{1} _{\Tilde{V}}} + 2 \sum_{k \in B_i} \dpd{A_{k*}} {\textbf{1}_{\Tilde{U}}} \Big) \\
                 &=& (1 + \frac{3\epsilon}{4})
                 \Big( \sum_{i \in \Tilde{L}}  \sum_{k \in B_i} \dpd{A_{k*}} {\textbf{1}_{\Tilde{V}}} + 2 \sum_{i \in \Tilde{L}} \sum_{k \in B_i} \dpd{A_{k*}} {\textbf{1}_{\Tilde{U}}} \Big) \\
                 &=&  (1 + \frac{3\epsilon}{4})
                 \Big(  \textbf{1}_{\Tilde{V}}^T A_{\Tilde{V}, \Tilde{V}} \textbf{1}_{\Tilde{V}} + 2 \textbf{1}_{\Tilde{V}}^T A_{\Tilde{V}, \Tilde{U}} \textbf{1}_{\Tilde{U}} \Big) \\
                 &\leq&  (1 + \frac{3\epsilon}{4})
                 \Big(  \textbf{1}^T A \textbf{1}  \Big)
\end{eqnarray*}
\end{itemize}
\end{proof}


Recall that we have assumed $m/6\leq \ell \leq m$. Under this assumption, the above claim says that $\widehat{m}$ is in fact a $(1\pm \eps)$-approximation to $m$. From the description of Algorithm~\ref{algo_onebfe}, the number of \iporacle{} queries made by the algorithm is $\widetilde{O}\left(\frac{\sqrt{\rho}n}{\sqrt{\ell}}\right)=\widetilde{O}\left(\frac{\sqrt{\rho}n}{\sqrt{{\bf 1}^T A {\bf 1}}}\right)$ as $m/10\leq \ell \leq m$. We will discuss how to remove the assumption, that $m/6\leq \ell \leq m$, by using a standard technique in property testing in the following.

\subsection{Proof of Theorem~\ref{theo:algo_quad}. How to remove the assumption that $\ell$ is a correct lower bound on $m=\textbf{1}^TA \textbf{1}$?}
\label{sec:removeL}
\remove{\begin{theo}[Theorem~\ref{theo:algo_quad} restated]\label{theo:bilin-dup}
There exists a query algorithm that for \onebfe that takes $\eps \in (0,1/2)$ as input and determines an $(1 \pm \eps)$-approximation to 
$\textbf{1}^TA \textbf{1}$ with high probability by using $\tOh\left(\frac{\sqrt{\rho} n}{\sqrt{\textbf{1}^TA \textbf{1}}}\right)$ many \iporacle{} queries to a symmetric matrix $A \in [[\rho]]^{n \times n}$.
Moreover, the algorithm only uses \iporacle{} query of the form $\dpd{A_{k*}}{{\bf u}}$ for some $k \in [n]$ and ${\bf u} \in \{0,1\}^n$. 
\end{theo}}
\begin{algorithm}
\caption{\onebfe($\epsilon$)}
\label{algo_onebfe_main}
\KwIn{$\epsilon \in (0,1)$.} 
\KwOut{$\widehat{m}$, which is an $(1 \pm \eps)$-approximation of
$\textbf{1}^TA \textbf{1}$.}
\Begin
	{
Initialize $\ell=\rho n^2/2$.

\While {($\ell \geq 1$)}
{
Call \onebfe($\ell,\epsilon$) (Algorithm~\ref{algo_onebfe}) and let $\widehat{m_\ell}$ be the output.

\If{$\left(\ell \leq \frac{\widehat{m_\ell}}{1+{3\eps}/{4}}\right)$}
{Report $\widehat{m_{\ell}}$ as the output and {\sc Quit}.}
\Else
{ 
Set $\ell$ as $\ell/2$ and {\sc Continue}.
}
}
Compute ${m}=\textbf{1}^TA \textbf{1}$ exactly, by making $n$ many \iporacle{} queries, and return it as $\widehat{m}$. 
}
\end{algorithm}
\remove{\begin{cl}[Claim~\ref{cl:approx} restated]
\label{cl:approx1}
With high probability, we have, (i) $ \widehat{m} \geq \left(1-\frac{\eps}{2}\right)\left(m-\frac{\eps}{4}\ell\right)$,  and (ii) $\widehat{m}\leq  \left(1+\frac{3\eps}{4}\right)m$.
\end{cl}}
  Algorithm~\ref{algo_onebfe_main} (\onebfe$(\eps)$) is our algorithm to determine an $(1\pm \eps)$ approximation to $m={\bf 1}^T A{\bf 1}$ with high probability. Note that \onebfe$(\eps)$ calls Algorithm~\ref{algo_onebfe} (\onebfe$(\ell,\eps)$ recursively at most $\Oh(\log (\rho n^2))$ times for different values of $\ell$ until we have $\ell \leq \frac{\widehat{m_\ell}}{1+{3\eps}/{4}}$. 
\begin{obs}\label{obs:obs1}
 \onebfe$(\eps)$ does not {\sc Quit} as long as $\ell>m$, with high probability.
\end{obs}
\begin{proof}
Consider a fix $\ell$ with $\ell>m$. By Claim~\ref{cl:approx1} (ii),  $\widehat{m_\ell} \leq \left(1+3\eps/4\right)m$ with high probability. As $\ell > m$, $\widehat{m_\ell} < \left(1+3\eps/4\right)\ell$ with high probability. This implies $\ell >\frac{\widehat{m_\ell}}{1+3\eps/4} $, that is, \onebfe$(\eps)$ does not {\sc Quit} for this fixed $\ell$ with high probability. As there can be at most $\Oh(\log(\rho n^2))$ many $\ell$'s with $\ell >m$ such that \onebfe$(\eps)$ calls \onebfe$(\ell,\eps)$, we are done with the proof. 
\end{proof}
\begin{obs}\label{obs:obs2}
Let $\ell \leq m/3$. \onebfe$(\eps)$ quits and reports $\widehat{m_{\ell}}$ as the output with high probability, where $\widehat{m_\ell}$ denotes the output of \onebfe$(\ell,\eps)$ 
\end{obs}
\begin{proof}
By Claim~\ref{cl:approx1}, with high probability, $\left(1-\frac{\eps}{2}\right)\left(m-\frac{\eps}{4}\ell\right)\leq \widehat{m}\leq  \left(1+\frac{3\eps}{4}\right)m$. As $\ell \leq m/3$, with high probability, we have $\left(1-\frac{3\eps}{4}\right)m \leq \widehat{m}\leq  \left(1+\frac{3\eps}{4}\right)m$. So, 
$\ell \leq \frac{\widehat{m_\ell}}{2(1-3\eps/4)}$ with high probability. As $\eps \in \left(0,\frac{1}{2}\right)$ (See the statement of Theorem~\ref{theo:bilin-dup}), we have  $\ell \leq \frac{\widehat{m_\ell}}{1+3\eps/4}$ with high probability. Hence, \onebfe$(\eps)$ quits and reports $\widehat{m_{\ell}}$ as the output with high probability.
\end{proof}
From Observation~\ref{obs:obs1} and~\ref{obs:obs2}, \onebfe$(\eps)$ does not quit when \onebfe$(\ell, \eps)$ is called for any $\ell >m$ and quits when \onebfe$(\ell, \eps)$ is called for some $ m/6\leq \ell \leq m$, with high probability. As \onebfe$(\eps)$ reduces $\ell$ by a factor of $2$ each time it does not quit, \onebfe$(\ell,\eps)$ is called $\Oh(\log(\rho n^2))$ times. The number of queries made by \onebfe$(\eps)$ for a call to \onebfe$(\ell,\eps)$ is $\tOh\left(\frac{\sqrt{\rho} n}{\sqrt{m}}\right)$. Observe that the query complexity of \onebfe$(\eps)$ is dominated by the query complexity of  last call to \onebfe$(\ell,\eps)$, that is, when $ m/6\leq \ell \leq m$. Hence, the total number of queries made by \onebfe$(\eps)$ is 
$\tOh\left(\frac{\sqrt{\rho} n}{ \sqrt{m}}\right)$.

%% file: sample.tex
\vspace{-2pt}
\subsection{Algorithm for Sample Almost Uniformly}
\label{sec:sample}
\vspace{-2pt}
\noindent
In this section, we will be proving the following theorem on almost uniformly sampling the entries of a symmetric matrix $A \in [[\rho]]^{n\times n}$. In Appendix~\ref{app:results}, we show how this algorithm can be extended to solve the more general \sau{A}{\bx}{\by} problem.

\begin{theo}\label{theo:algo_samp}
Let $A \in [[\rho]]^{n\times n}$ be an unknown symmetric matrix with \iporacle{} query access. 
There exists an algorithm that takes $\eps \in (0,1)$ as input and with high probability outputs a sample from a distribution on $[n]\times [n]$, such that each $(i,j) \in [n]\times [n]$ is sampled with probability $p_{ij}$ satisfying:
$(1-\eps)\frac{A_{ij}}{\sum\limits_{1\leq i, j \leq n} A_{ij}} \leq p_{ij} \leq (1+\eps)\frac{A_{ij}}{\sum\limits_{1\leq i, j \leq n} A_{ij}}$.

Moreover, the algorithm makes $\tOh\left(\frac{\sqrt{\rho} n}{\sqrt{\oneaone}}\right)$ \iporacle{} queries to the matrix $A$ of the form $\dpd{A_{k*}}{{\bf{u}}}$ for some $k \in [n]$ and ${\bf u} \in \{0,1\}^{n}$.
\end{theo}

Our algorithm for {\sc Sample Almost Uniformly} is a generalization of Eden and Rosenbaum's algorithm for \emph{sampling edges of an unweighted graph}~\cite{EdenR18}. First, consider the following strategy by which we sample each ordered pair $(i,j)\in [n]\times[n]$ proportional to $A_{ij}$ when the matrix $A$ is such that $\dpd{A_{i,*}}{{\bf 1}}$ is the same for each 
$i \in [n]$.

\noindent\underline{{\em Strategy-1:}} Sample $r \in [n]$ uniformly at random and then sample an ordered pair of the form $(r,j)$ from the $r$-th row using \regr{} query.
Observe that this strategy fails when $\dpd{A_{i*}}{{\bf 1}}$'s are not the same for every $i \in [n]$. So, the modified strategy is as follows.

\noindent\underline{{\em Strategy-2:}} Sample $r \in [n]$ with probability $\frac{\dpd{A_{r*}}{{\bf 1}}}{{\bf 1}^T A {\bf 1}}$ and then sample an ordered pair of the form $(r,j)$ from the $r$-th row by using \regr{} query.

Note that Strategy-2 samples each ordered pair $(i,j)$ proportional to $A_{ij}$. However, there are two challenges in executing Strategy-2:
\begin{itemize}
\item[(i)] We do not know the value of ${\bf 1}^T A {\bf 1}$.
\item[(ii)] We need $\Omega(n)$ queries to determine $\dpd{A_{r*}}{{\bf 1}}$ {for each $r \in [n]$ for all $r \in [n]$}.
\end{itemize}
 The first challenge can be taken care of by finding an estimate $\hat{m}$ for ${\bf 1}^T A {\bf 1}$, with high probability, by using Theorem~\ref{theo:algo_quad} such that {$\hat{m} =\Theta( {\bf 1}^T A {\bf 1})$}. To cope up with the second challenge, we partition the elements as well as rows into two classes as defined in Definition~\ref{defi:light_heavy_row}. In what follows, we consider a parameter $\tau$ in terms of which we base our discussion as well as algorithm. $\tau$ is a function of $\hat{m}$ that will evolve over the calculation and will be $\tau=\sqrt{\frac{\rho \hat{m}}{\eps}}$.\begin{defi}\label{defi:light_heavy_row}
 The $i$-th row of the matrix is light if $\dpd{A_{i,*}}{\bf{\bf{1}}} $ is at most $\ths$. Otherwise, the $i$-th row is heavy. Any order pair $(i,j)$, for a fixed $i$, is light (heavy) if the $i$-th row is light (heavy).
\end{defi}
We denote the set of all light (heavy) ordered pairs by $\cL$ ($\cH$). Also, let $I(L)$ ($I(H)$) denote the set of light (heavy) rows of the matrix $A$. Let $w(\cL)=\sum_{A_{ij} \in \cL} A_{ij}$ and $w(\cH)=\sum_{A_{ij} \in \cH}  A_{ij}$.

Our algorithm consists of repeated invocation of two subroutines, that is, \samplelight and \heavy. Both \samplelight and \heavy succeed with \emph{good} probability and sample elements from $\cL$ and $\cH$ almost uniformly, respectively. The threshold $\ths$ is set in such a way that there are \emph{large}~\footnote{Large is parameterized by $\ths$.} number of light rows and small number of heavy rows. In \samplelight, we select a row uniformly at random, and if the selected row is light, then we sample an ordered pair from the selected row randomly using \regr. This gives us an element from $\cL$ uniformly. However, the same technique will not work for \heavy as we have few heavy rows. To cope up with this problem, we take a row uniformly at random and if the selected row is light, we sample an ordered pair from the selected row randomly using \regr{}. Let $(i,j)$ be the output of the \regr{} query. Then we go to the $j$-th row, if it is heavy, and then select an ordered pair from the $j$-th row randomly using \regr{} query.

 The formal algorithms for \samplelight and \heavy are given in Algorithm~\ref{algo_light} and Algorithm~\ref{algo_heavy}, respectively. 
 The formal correctness proofs of \samplelight and \heavy are given in 
 Lemmas~\ref{lem:sample-light} and \ref{lem:heavy}, respectively. We give the final algorithm along with its proof of correctness in Theorem~\ref{theo:algo_samp}.

\begin{algorithm}
\caption{\samplelight}
\label{algo_light}
\KwIn{An estimate $\hat{m}$ for ${\bf 1}^T A {\bf 1}$ and a threshold $\ths$.}
\KwOut{$(i,j) \in \cL$ with probability $\frac{A_{ij}}{n \ths}$.}
\Begin
	{
	Select a row $r \in [n]$ uniformly at random.\\
	\If{($r \in I(\cL)$, that is, $\dpd{A_{r*}}{\bf{\bf{1}}}$ is at most $ \ths$)}
	{
	Return {\sc Fail} with probability $p=\frac{\tau - \dpd{A_{r*}}{\bf{\bf{1}}} }{\tau}$ , and 
	Return \regr$(r,\bf{1})$ with probability $1-p$ as the output.
	}
	
	{ Return {\sc Fail}}
}
\end{algorithm}

\begin{lem}\label{lem:sample-light}
\samplelight succeeds with probability $\frac{w(\cL)}{n \tau}$.  Let $Z_{\ell}$ be the output in case it succeeds. Then $\pr(Z_{\ell}=(i,j))  = \frac{A_{ij}}{n \ths}$ if $(i,j) \in \cL$, and $\pr(Z_{\ell}=(i,j) ) = 0$, otherwise. Moreover, \samplelight makes $\Oh(\log n)$ many queries.
\end{lem}
\begin{proof}
Consider an ordered pair $(i,j) \in \cL$. The probability of $(i,j)$ returned by \samplelight is 
\begin{eqnarray*}
&&\pr(Z_{\ell}=(i,j))\\
&&= \pr(r=i) \cdot \pr(\mbox{\samplelight returns \regr$(r,\bf{1})$})\cdot \pr(\mbox{\regr$(r,\bf{1})$~returns}~ (i,j))\\
&&= \frac{1}{n}\cdot \frac{\dpd{A_{r*}}{\bf{\bf{1}}}}{\tau} \cdot \frac{A_{ij}}{\dpd{A_{r*}}{\bf{\bf{1}}}}=\frac{A_{ij}}{n \tau}
\end{eqnarray*}
Hence, the probability that \samplelight does not return {\sc Fail} is $\sum_{A_{ij} \in \cL} \frac{A_{ij}}{n \tau} = \frac{w(\cL)}{n \tau}$.
Now the query complexity of \samplelight follows from the query complexity of \regr{} given in Lemma~\ref{lem:random}.
\end{proof}

\begin{lem}\label{lem:heavy}
\heavy succeeds with probability at most $\frac{w(\cH)}{n \ths}$ and at least $\left( 1-\frac{\rho \hat{m}}{\ths^2}\right)\frac{w(\cH)}{n \ths}$. Let $Z_{h}$ be the output in case it succeeds. Then $\left( 1-\frac{\rho \hat{m}}{\ths^2}\right)\frac{A_{ij}}{n \ths} \leq \pr(Z_h =(i,j)) \leq\frac{A_{ij}}{n \ths}$ for each 
$(i,j) \in \cH$, and $\pr(Z_h =(i,j))=0$, otherwise. Moreover, \heavy makes $\Oh(\log n)$ many queries.
\end{lem}
\begin{proof}
For each $k \in I(\cH)$, note that, $\dpd{A_{k*}}{\bf{1}}$ is more than $\ths$. So, $$
    \size{I(\cH)} \leq \frac{\oneaone}{\ths} \leq  \frac{\hat{m}}{\ths}.
$$
Note that
$$
    \dpd{A_{k*}}{{\vecone}} = \sum_{u \in I(\cL)} A_{ku} + \sum_{v \in I(\cH) } A_{kv}.
$$
Observe that 
$$
    \sum_{v \in I(\cH) } A_{kv} \leq \rho \size{I(\cH)} \leq \frac{\rho \hat{m}}{\ths} \leq \frac{\rho \hat{m} \dpd{A_{k*}}{\bf{1}}}{\ths^2}.
$$
So, we have the following Observation.
\begin{obs}\label{obs:inter}
 $ \sum_{u \in I(\cL)} A_{k u}\geq \left( 1- \frac{\rho \hat{m}}{\ths^2}\right) \dpd{A_{k*}}{\bf{1}}$, where $k \in I(\cH)$.
\end{obs}

Let us consider some ordered pair $(i,j) \in \cH$. The probability that $(i,j)$ is returned by the algorithm is 
\begin{eqnarray*}
\pr(Z_h=(i,j)) &=& \pr(s=i) \cdot \pr(\mbox{\regr$(s,\bf{1})$ returns $(i,j)$})\\
&=& \left(\sum\limits_{u \in I(\cL)} \pr(r=u) \cdot \frac{\dpd{A_{r*}}{\bf{1}}}{\ths} \cdot \pr (\mbox{\regr$(r,\bf{1})$ returns $(r,i)$}) \right) \cdot \frac{A_{ij}}{\dpd{A_{i*}}{\bf{1}}} \\
&=& \frac{1}{n} \cdot\frac{A_{ij}}{\dpd{A_{i*}}{\bf{1}}} \cdot \sum\limits_{u \in I(\cL)} \frac{\dpd{A_{u *}}{\bf{1}}}{\ths} \cdot \frac{A_{u i}}{\dpd{A_{u*}}{\bf{1}}}= \frac{A_{ij}}{n\ths}   \cdot \frac{\sum\limits_{u \in I(\cL)} {A_{i u}}}{\dpd{A_{i*}}{\bf{1}}}
\end{eqnarray*}
The last equality follows from the fact that 
$A$ is a symmetric matrix. 

Using the fact that $\sum\limits_{u \in I(\cL)} {A_{i u}} \leq \dpd{A_{i*}}{\bf{1}}$ and Observation~\ref{obs:inter}, we have 
$$
    1-\frac{\rho\hat{m}}{\ths^2}   \leq \frac{1}{\dpd{A_{i*}}{\bf{1}}}\, \sum\limits_{u \in I(\cL)} {A_{i u}} \leq 1.
$$
Putting everything together, we get
$$
    \left(1-\frac{\rho\hat{m}}{\ths^2}\right) \cdot \frac{A_{ij}}{n \ths} \leq \pr(Z_h=(i,j)) \leq \frac{A_{ij}}{n \ths}.
$$
 
So, the probability that \heavy succeeds is $\sum\limits_{A_{ij} \in \cH} \pr(Z=(i,j))$, which lies between $ \left(1-\frac{\rho \hat{m}}{\ths^2}\right) \frac{w(\cH)}{n \ths}$ and $\frac{w(\cH)}{n \ths}$.
The query complexity of \heavy follows from the query complexity of \regr{} (Lemma~\ref{lem:random}).
\end{proof}

\begin{algorithm}
\caption{ \heavy($\hat{m}$)}
\label{algo_heavy}
\KwIn{An estimate $\hat{m}$ for ${\bf 1}^T A {\bf 1}$ and a thershold $\ths$.}
\KwOut{$A_{ij} \in \cH$ with probability at most $\frac{A_{ij}}{n \ths}$ and at least $\left( 1-\frac{\rho \hat{m}}{\ths^2}\right)\frac{A_{ij}}{n \ths}$.}
\Begin
	{
	Select a row $r \in [n]$ uniformly at random\;
	
	\If{($r \in I(\cL)$, that is, $\dpd{A_{r*}}{\bf{\bf{1}}}$ is at most $ \ths$)}
	{
	Return {\sc Fail} with probability $p=\frac{\tau - \dpd{A_{r*}}{\bf{\bf{1}}} }{\tau}$, and with probability $1-p$ do the following\;
	 $A_{rs} = $\regr($r,\bf{1}$)\\
	 If $s\in I(\cH)$, that is, $\dpd{A_{s*}}{\bf{\bf{1}}}>\ths$, then
	 	Return \regr($s,\bf{1}$) as the output. Otherwise, 	{Return {\sc fail}\;
	 }
	}
	
	{Return {\sc Fail}\;
	}
}
\end{algorithm} 
Now we will prove Theorem~\ref{theo:algo_samp}.
\begin{proof}[Proof of Theorem~\ref{theo:algo_samp}.]
Our algorithm first finds a rough estimate $\hat{m}$ for ${\bf 1}^T A {\bf 1}$, with high probability, by using Theorem~\ref{theo:algo_quad} such that {$ \hat{m} = \Theta( {\bf 1}^T A {\bf 1})$}. For the rest of the proof, we work on the conditional probability space that {$ \hat{m} = \Theta( {\bf 1}^T A {\bf 1})$}. We set $\ths = \sqrt{\frac{ \rho \hat{m}}{\eps}}$ and do the following for $\Gamma$ times, where  $\Gamma$ is a parameter to be set later. With probability $1/2$, we invoke \samplelight and with probability $1/2$, we invoke \heavy.
If the ordered pair $(i,j)$ is reported as the output by either \samplelight or \heavy, we report that. If we get {\sc Fail} in all the trials, we report {\sc Fail}.

Now, let us consider a particular trial and compute the probability of success $\pr(S)$, which is
$\pr(S) = \frac{1}{2}\left(\pr(\mbox{\samplelight  succeeds)}+\pr(\mbox{\heavy  succeeds}) \right)$.
Observe that from Lemmas~\ref{lem:sample-light} and~\ref{lem:heavy}, we have,
$\frac{1}{2}\left( \frac{w(\cL)}{n \ths} + \left(1-\frac{\rho \hat{m}}{\ths^2} \right) \frac{w(\cH)}{n \ths}\right) \leq \pr(S) \leq \frac{1}{2} \left( \frac{w(\cL)}{n \ths} + \frac{w(\cH)}{n \ths} \right)$. This implies $(1-\eps) \frac{\oneaone}{2n \ths} \leq \pr(S) \leq \frac{\oneaone}{2n \ths}$ as $\ths=\sqrt{\frac{ \rho \hat{m}}{\eps}}$ and using $w(\cL)+w(\cH)=\oneaone$.


Now, let us compute the probability of the event $\cE_{ij}$, that is, the algorithm succeeds and it returns $A_{ij}$. If $A_{ij} \in \cL$, by Lemma~\ref{lem:sample-light}, we have $\pr(Z=(i,j)) = \frac{1}{2} \cdot \frac{A_{ij}}{n \ths} $. Also, if $A_{ij} \in \cH$, by Lemma~\ref{lem:heavy}, we have,
$\left( 1-\frac{\rho \hat{m}}{\ths^2}\right)  \frac{A_{ij}}{2 n \ths} \leq \pr(Z =(i,j)) \leq \frac{A_{ij}}{2 n \ths}$.
So, for any $(i,j)$, we get
$\left( 1-\eps \right)\frac{A_{ij}}{2n \ths} \leq \pr(\cE_{ij}) \leq  \frac{A_{ij}}{2n \ths}$.
Let us compute the probability of $\cE_{ij}$ on the conditional probability space that the algorithm succeeds, that is, 
$ \pr(Z=(i,j)~|~S) = \frac{\pr(\cE_{ij})}{\pr(S)}$, which lies in the interval $ \left[(1-\eps)\frac{A_{ij}}{\oneaone}, (1+\eps)\frac{A_{ij}}{\oneaone}\right]$ as $\eps \in \left(0,\frac{1}{2} \right)$.

To boost the probability of success, we set $\Gamma= \Oh\left( \frac{n\sqrt{\rho}}{(1-\eps) \sqrt{\eps \hat{m}}}\log n\right)$ for a suitable \emph{large} constant in $\Oh(\cdot)$ notation.
The query complexity of each call to \samplelight and \heavy is $\Oh(\log n)$. Also note that our algorithm for \onealmostuni makes at most $ \Oh\left( \frac{n\sqrt{\rho}}{(1-\eps) \sqrt{\eps \hat{m}}}\log n\right)$ many invocation to \samplelight and \heavy.\remove{ So, the query complexity of each iteration is $\Oh(\log n)$ and our algorithm has $ \Oh\left( \frac{n\sqrt{\rho}}{(1-\eps) \sqrt{\eps \hat{m}}}\log n\right)$ iteration.} Hence, the total query complexity of our algorithm is $\tOh\left(\frac{\sqrt{\rho} n}{\sqrt{\oneaone}}\right)$.
\end{proof}

%% file: more-results.tex
\remove{
 We first discuss our results when $A$ is a symmetric matrix, $A_{ij} \in [[\rho]]~\forall i,j \in [n]$ and $\bf{x}=\bf{y}=\vecone$. Though our main focus is to estimate the bilinear form $ \bx^TA\by $, we show that it is enough to consider the case of symmetric matrix $A$ and $\bf{x} = \bf{y} = 1$, that is, $\oneaone$. Then we show how our results for the special case can be extended to the general case.

 \begin{table}[h!]
\centering
\begin{tabular}{||c | c | c||} 
 \hline
 Problem & Upper bound & Lower bound\\
 \hline\hline
 \bfe{\{0,1\}^n}{\vecone}{\vecone} & $\tOh\left(\frac{\sqrt{\rho}n}{\sqrt{\oneaone}}\right)$ & $\Omega\left(\frac{\sqrt{\rho}n}{\sqrt{\oneaone}}\right)$ \\
\hline
\sau{\R^n}{\vecone}{\vecone} & $\tOh\left(\frac{\sqrt{\rho} n}{\sqrt{\oneaone}}\right)$ & $\Omega\left(\frac{\sqrt{\rho}n}{\sqrt{\oneaone}}\right)$ \\
 \hline
\end{tabular}
\caption{Query complexities for \bfe{S}{\bx}{\by} and \sau{S}{\bx}{\by} when $\bx=\by=\vecone$. The upper bound holds for $S=\{0,1\}^n$ and the lower bound holds for $S=\R^n$. The stated result is for any matrix $A$ and when $A_{ij} \in [[\rho]]~\forall i,j \in [n]$.}
\label{table:graphresults}
\end{table}

\paragraph*{$A$ is symmetric and $\bx=\by=\vecone$.}
In this case we have the results mentioned in Table~\ref{table:graphresults}.\remove{ and the results are formally stated in Theorem~\ref{} and~\ref{}.}
Note that the queries made in the above case when $A$ is symmetric are only row \iporacle{\{0,1\}^n} queries.
}

\noindent
We will first discuss the extensions of our results, from the previous sections, to solve $\onebfe_{A}(\vecone,\vecone)$ and $\onesau_{A}(\vecone,\vecone)$ when $A\in [[\rho]]^{n\times n}$ is not necessarily a symmetric matrix. Consider the symmetric matrix $B =\frac{A+A^T}{2}$. Using Theorem~\ref{theo:algo_quad}, we can solve $\onebfe_{B}(\vecone,\vecone)$ and $\onesau_{B}(\vecone,\vecone)$ only using row \iporacle{} queries on the matrix $B$. Observe that one can simulate the row \iporacle{} access for $B$ by using two 
\iporacle{} queries to $A$ as $\dpd{B_{i*}}{\bf{v}}=\frac{\dpd{A_{i*}}{\bf{v}}+\dpd{A_{*i}}
{\bf{v}}}{2}$. 
So, we can solve both $\onebfe_{B}(\vecone,\vecone)$ and $\onesau_{B}(\vecone,\vecone)$ using $\tOh\left(\frac{\sqrt{\rho}n}{\sqrt{{\bf{x}}^{T}B{\bf{y}}}}\right)$ many row and column \iporacle{}
queries to the matrix $A$. Note that the query algorithms for both $\onebfe_{B}(\vecone,\vecone)$ and $\onesau_{B}(\vecone,\vecone)$ problems uses queries of the form $\dpd{A_{i*}}{\bf{v}}$ or $\dpd{A_{*i}}{\bf{v}}$ where $i \in [n]$ and $\bf{v} \in \{0,1\}^{n}$.

Now, let us return to the problems $\onebfe_{A}(\vecone,\vecone)$ and $\onesau_{A}(\vecone,\vecone)$ that we wanted to solve.
We will now convert the query algorithms for matrix $B$ to algorithms for the matrix $A$. 
\begin{itemize}
    \item
        As ${\bf{1}}^{T} A {\bf{1}} = {\bf{1}}^{T} B {\bf{1}}$, we can directly use the query algorithm for $\onebfe_{B}(\vecone, \vecone)$ for the problem $\onebfe_{A}(\vecone, \vecone)$. 

    \item
        Now consider the problem $\onesau_{A}(\vecone, \vecone)$. We will first run the query algorithm for $\onesau_{B}(\vecone, \vecone)$.
        Suppose the algorithm outputs $(i,j) \in [n]\times [n]$, which happens with probability $p_{ij}$ where $p_{ij} \in\left[(1-\eps)
        \frac{B_{ij}}{{\vecone}^T B {\vecone}}, (1+\eps)\frac{B_{ij}}{{\vecone}^T B {\vecone}}\right]$. Now we want to use the fact that $B_{ij} = \frac{A_{ij}+A_{ji}}{2}$. We find $A_{ij}$ and $A_{ji}$ by two row \iporacle{} queries to 
        the matrix $A$ as $A_{ij}=\dpd{A_{i*}}{\bf{a_j}}$ and $A_{ji}=\dpd{A_{j*}}{\bf{a_i}}$, where $ \bf{a_i}$ ($
        \bf{a_j}$) is the vector in $\{0,1\}^n$ having $1$ only in the $i$-th ($j$-th) entry. We report $(i,j)$ 
        and $(j,i)$ with probability $\frac{A_{ij}}{A_{ij}+{A_{ij}}}$ and $\frac{A_{ij}}{A_{ij}+{A_{ij}}}$, 
        respectively. Observe that $A_{ij}$ can be the output only when $B_{ij}$ or $B_{ji}$ is the output of the query algorithm for the problem $\onesau_{B}(\vecone,\vecone)$.
        Using the facts that $B_{ji}=B_{ij}=\frac{A_{ij} + A_{ji}}{2}$ and $\oneaone = {{\vecone}^T B 
        {\vecone} }$, we report $A_{ij}$ with probability in the range $\left[(1-\eps)\frac{A_{ij}}{{\vecone}^T A {\vecone}}, (1+\eps)
        \frac{A_{ij}}{{\vecone}^T A {\vecone}}\right]$.
        \remove{
        Hence, the results mentioned in Table~\ref{table:graphresults} also hold for the case when the matrix $A$ is not necessarily symmetric.}
        
    \item
        Therefore 
        the query algorithms for $\onebfe_{A}(\vecone,\vecone)$ and $\onesau_{A}(\vecone,\vecone)$ uses $\tOh\left(\frac{\sqrt{\rho}n}{\sqrt{{\bf{x}}^{T}B{\bf{y}}}}\right) = \tOh\left(\frac{\sqrt{\rho}n}{\sqrt{{\bf{x}}^{T}A{\bf{y}}}}\right)$ (as ${\bf{1}}^{T} A {\bf{1}} = {\bf{1}}^{T} B {\bf{1}}$) many row and column \iporacle{} queries to the matrix $A$. 
\end{itemize}

Finally, we now consider the general problems \bfe{A}{\bx}{\by} and \sau{A}{\bx}{\by} where $A \in [[\rho]]^{n\times n}$ (not necessarily symmetric), $\bf{x} \in [[\gamma_{1}]]$ and $\bf{y} \in [[\gamma_{2}]]$.
Observe that ${\vecone}^T C {\vecone} = {{\bf{x}}^T A {\bf{y}}}$, where $C_{ij} = A_{ij} x_{i} y_{j}$. So, if we have
\iporacle{} oracle access to the matrix $C$, then we can design query algorithms for $\onebfe_{A}(\bf{x}, \bf{y})$ and $\onesau_{A}(\bf{x}, \bf{y})$ problems by using the query algorithms for $\onebfe_{C}(\vecone, \vecone)$ and $\onebfe_{C}(\vecone, \vecone)$ respectively. Note that, as we have already discussed, that even if $C$ is not symmetric we can still design efficient query algorithms for $\onebfe_{C}(\vecone, \vecone)$ and $\onebfe_{C}(\vecone, \vecone)$. 

But, we do not directly have \iporacle{} query access to $C$. However, we can simulate any \iporacle{} query to $C$ by using an appropriate \iporacle{} query to $A$ by the following observation.
\begin{obs}\label{obs-IP-C-A}
 $\dpd{C_{k*}}{{\bf{a}}}= \dpd{A_{k*}}{\bf{a'}}$, where ${a'}_i=x_k{a}_iy_i$ for each $i \in [n]$. And  $\dpd{C_{*k}}{{{a}}}= \dpd{A_{*k}}{{a'}}$, where ${a'}_i=x_i{a}_iy_k$ for each $i \in [n]$.
\end{obs}

\remove{
For the query complexity, observe that we are solving ${\bf{1}^T C {\bf{1}}}$ and $C_{ij} \in [\rho \gamma_1 \gamma_2]$. Now using the result mentioned in Table~\ref{table:graphresults}, we have the following results mentioned in Table~\ref{table:graphresults1} for \bfe{S}{\bx}{\by} and \sau{S}{\bx}{\by}.

\begin{table}[h!]
\centering
\begin{tabular}{||c | c | c||} 
 \hline
 Problem & Upper bound & Lower bound\\
 \hline\hline
 \bfe{A}{\bx}{\by} & $\tOh\left(\frac{\sqrt{\rho \gamma_1 \gamma_2}\left(\mathrm{nnz}(x) + \mathrm{nnz}(y) \right)}{\sqrt{{\bx}^{T} A {\by}}}\right)$ & $\Omega\left(\frac{\sqrt{\rho \gamma_1 \gamma_2}\left(\mathrm{nnz}(x) + \mathrm{nnz}(y) \right)}{\sqrt{{\bx}^{T} A {\by}}}\right)$ \\
\hline
\sau{\{0,1\}^n}{\bx}{\by} & $\tOh\left(\frac{\sqrt{\rho \gamma_1 \gamma_2} \left(\mathrm{nnz}(x) + \mathrm{nnz}(y) \right)}{\sqrt{{\bx}^{T} A {\by}}}\right)$ & $\Omega\left(\frac{\sqrt{\rho \gamma_1 \gamma_2}\left(\mathrm{nnz}(x) + \mathrm{nnz}(y) \right)}{\sqrt{{\bx}^{T} A {\by}}}\right)$ \\
 \hline
\end{tabular}
\caption{Query complexities for \bfe{S}{\bx}{\by} and \sau{S}{\bx}{\by}.
}
\label{table:graphresults1}
\end{table}
}

Using Observation~\ref{obs-IP-C-A} (i.e., \iporacle{} query to $C$ can be simulated by \iporacle{} query to $A$), Theorems~\ref{theo:algo_quad} and \ref{theo:algo_samp}, and the fact that $C_{ij} \in [[\rho\gamma_{1}\gamma_{2}]]$, we get the following result.  
\begin{theo}\label{theorem-A-not-symmetric-bilinearform-sampling-algo}
        Let $A \in [[\rho]]^{n\times n}$ be an unknown matrix (not necessarily symmetric) with an \iporacle{} query access. 
        There exist query algorithms for \bfe{A}{\bx}{\by} and \sau{A}{\bx}{\by} that takes $\bx \in [[\gamma_1]]^n$, $\by \in [[\gamma_2]]^n$ and $\eps \in (0,1/2)$ as inputs, and gives the { correct output} with probability at least $2/3$, using $\tOh\left(\frac{\sqrt{\gamma_1 \gamma_2\rho} \left(\mathrm{nnz}(x) + \mathrm{nnz}(y)\right)}{\sqrt{\textbf{x}^TA \textbf{y}}}\right)$ many \iporacle{} queries to $A$.
        Additionally, if ${\bf x}, \, {\bf y} \in \{0,1\}^{n}$, the query algorithms only uses queries of the form $\dpd{A_{k*}}{{\bf{u}}}$
        or $\dpd{A_{*k}}{{\bf{u}}}$ for some $k \in [n]$ and ${\bf u} \in \{0,1\}^{n}$.
\end{theo}
We also show the following lower bound that matches the upper bound of 
Theorem~\ref{theorem-A-not-symmetric-bilinearform-sampling-algo}.
\begin{theo}
\label{theorem-A-not-symmetric-bilinearform-sampling-lower-bound}
            Let $A \in [[\rho]]^{n\times n}$ be an unknown matrix with an \iporacle{} query access. 
        Any algorithm for \bfe{A}{\bx}{\by} and \sau{A}{\bx}{\by} that takes $\bx \in [[\gamma_1]]^n$, $\by \in [[\gamma_2]]^n$ and $\eps \in (0,1/2)$ as inputs, and gives the correct output with probability at least $2/3$, will require $\Omega\left(\frac{\sqrt{\rho \gamma_1 \gamma_2} \left(\mathrm{nnz}({\bf x}) + \mathrm{nnz}({\bf y})\right)}{\sqrt{\textbf{x}^TA \textbf{y}}}\right)$ many \iporacle{} queries to $A$.
\end{theo}
\begin{proof}
We show that the stated lower bound for \bfe{A}{\bx}{\by} holds even if (i) $A$ is a symmetric matrix, (ii) $x_i=\gamma_1$ for each $i \in [n]$, that is, $\mathrm{nnz}({\bf x})=n$, and (iii) $y_i=\gamma_2$ for each $i \in [n]$, that is, $\mathrm{nnz}({\bf y})=n$. Without loss of generality assume that $m =o(\rho \gamma_1 \gamma_2n^2)$.

We prove by giving a reduction from {\sc Disjointness} in communication complexity: Alice is given $\boldvec{a} \in \{0,1\}^t$ and Bob is given $\boldvec{b} \in \{0,1\}^t$, where $t =
\frac{\sqrt{\rho \gamma_1 \gamma_2}n}{\sqrt{m}}-1$. The players want to output $1$ if there is an $i \in [t]$ such that $a_i = b_{i} = 1$. {\sc Disjointness} admits a randomized communication complexity of $\Omega(t)$ (See Definition~\ref{defi:k-intersection} and Lemma~\ref{theo:kinter}). Consider a block-diagonal matrix $A \in [[\rho]]^{n \times n}$ with $A^1, \dots, A^{t},A^{t+1} \in [[\rho]]^{K \times K}$ diagonal blocks. For $i \in [t]$, if $a_i = b_{i} = 1$, then $(A^i)_{(r,s)} = \rho$ for all $r,s \in [K]$; and if $a_i = b_{i} \neq 1$, then $A^i = 0$. Also, $(A^{t+1})_{(r,s)} = \rho$ for all $r,s \in [K]$.  Observe that, for $\boldvec{x} = \gamma_1^n$ and $\boldvec{y} = \gamma_2^n$, $\boldvec{x}^T A \boldvec{y} \geq 2m$ if $\boldvec{a}$  $\boldvec{b}$ intersect. Otherwise, $\boldvec{x}^T A \boldvec{y} = m$.
    
We will be done with the stated lower bound proof for \bfe{A}{\bx}{\by} by  showing that Alice and Bob can determine the answer to any \iporacle{} query, to matrix $A$, with $2$ bits of communication. Let $\dpd{A_{i*}}{\boldvec{v}}$ be an \iporacle{} query. From the construction of matrix $A$, there exists a block diagonal matrix $A^j, j \in [t+1]$, such that row $A_{i*}$ can completely determined if we know $A^j$. If $j=t+1$, then Alice and Bob need not communicate as it is known that $(A^{t+1})_{(r,s)} = \rho$ for all $r,s \in [K]$. If $1\leq j\leq t$, $A^j$ depends on $x_j$ and $y_j$. So, Alice and Bob can determine matrix $A^j$, and hence, $\dpd{A_{i*}}{\boldvec{v}}$ with $2$ bits of communication. Similar argument holds for any \iporacle{} query of the form $\dpd{A_{*i}}{\boldvec{v}}$.

Recall the argument that we used to prove the lower bound for {\sc Induced Edge Sampling} (Theorem~\ref{theo-lower-bound-induced-sampling-local-resate}) by using the construction of the lower bound proof of {\sc Induced Edge Esrimation} (Theorem~\ref{theo-lower-bound-induced--RESTATE}). One can use similar argument to show
the stated lower bound for \sau{A}{\bx}{\by} by using the construction of our lower bound proof for \bfe{A}{\bx}{\by}. 
\end{proof}

%% file: conclude1.tex
\section{Conclusion and discussions}
\label{sec:conclude}

\subsection{Other Matrix Problems}
\label{app:matrix}

\noindent {Recently, vector-matrix query~\cite{SWYZ19} and vector-matrix-vector query~\cite{DBLP:conf/approx/RashtchianWZ20} were introduced to study a bunch of matrix, graph and statistics problems. As noted earlier,  \iporacle{} query oracle is in the same linear algebraic framework of vector-matrix query and vector-matrix-vector query, but these queries are stronger than \iporacle{} query. Study of the various matrix, graph and statistics problems, introduced in~\cite{SWYZ19,DBLP:conf/approx/RashtchianWZ20}, using \iporacle{} query will be of independent interest. As a first step in that direction, in Appendix~\ref{app:matrix}, we study the query complexity of the following problems using \iporacle{} queries.}
\begin{itemize}
    \item {\sc Symmetric Matrix}: Is $A\in \{0,1\}^{n\times n}$ a symmetric matrix? 
    
    \item {\sc Diagonal Matrix}: Is $A$ a diagonal matrix? 
    
    \item {\sc Trace}: Compute the trace of the matrix $A$.
    
    \item {\sc Permutation Matrix}: Is $A \in \{0,1\}^{n\times n}$ a permutation matrix?
    
    \item {\sc Doubly stochastic matrix}: Is $A \in \{0,1\}^{n\times n}$ a doubly stochastic matrix?
    
    \item {\sc Identical columns}: Does there exist two columns in $A\in \{0,1\}^{n\times n}$ that are identical?
    
    \item {\sc All ones column}: Does there exist a column in $A$ all of whose entries are $1$?
\end{itemize}
\remove{
\begin{table}[h]
\centering
\caption{Other matrix problems.}

\begin{tabular}{||c | c |c ||} 
 \hline
 {\bf Problem} & {\bf Query} & {\bf Comments}\\
 \hline\hline
  {\sc Symmetric Matrix} & $\widetilde{\Theta}\left(n\right)$ & Theorem~\ref{theo:ub_symm} \\ 
 \hline
 {\sc  Diagonal Matrix} &$\Theta\left(n\right)$ & Theorem~\ref{theo:ub_dia} \\ \hline
 {\sc Trace } & $\Theta\left(n\right)$& Theorem~\ref{theo:ub_trace} \\ \hline
  {\sc Permutation Matrix} &${\Theta}\left(n\right)$ & Theorem~\ref{theo:ub_perm}\\ \hline
  {\sc Doubly Stochastic matrix } &${\Theta}\left(n\right)$ &Theorem~\ref{theo:ub_sto} \\ \hline
    {\sc  Identical columns} & $\widetilde{\Theta}\left(n\right)$ & Theorem~\ref{theo:ub_identical} \\ \hline
  {\sc All Ones Columns} & ${\Theta}\left(n\right)$ & Theorem~\ref{theo:ub_allone}\\ \hline
\end{tabular}
\label{table:matrix-main}

\end{table} 
}

Table~\ref{table:matrix-main-app} (in Appendix~\ref{app:matrix}) gives the query complexity of solving the above matrix problems using IP oracle. {We also do a comparative study of \iporacle{} with respect to these stronger queries.} 

\subsection{Data structure complexity and open problems}
\noindent{\bf Data structure complexity:} Besides property testing, there have been extensive work concerning vector-matrix-vector product in data structure complexity and other models of computation like the cell probe model~\cite{DBLP:conf/stoc/CKL18, DBLP:conf/stoc/ChKL18, DellLM20, DBLP:conf/soda/LW17, DBLP:conf/innovations/RR20}. For the purposes of this paper, it is an interesting question to find a pre-processing scheme for the matrix such the \iporacle{} queries on the matrix can be answered efficiently.

\vspace{5pt}
\noindent{\bf Open questions:} {One of the open problem (that is left from our work) is to design algorithm and/or prove lower bound for {\sc Bilinear Form Estimation} and {\sc Sampling} when the entries of the matrices are not necessarily positive. Here we would like to that our technique does not work for matrix with both positive and negative entries.}

Some other natural open question are
\begin{itemize}
    \item Are there some special kind of matrices where we can solve {\sc Bilinear Form Estimation} and {\sc Sampling} using fewer queries?
    \item Can we solve some other linear algebraic problems using \iporacle{} queries?
    \item Are there other graph problems, where {\sc induced degree} outperforms {\sc local} queries?
\end{itemize}

%% file: conclude.tex
\section{Weighted edge estimation and weighted edge sampling}\label{sec:weighted} 
As an application, consider the {\sc Weighted Edge Estimation} problem on a graph $G$, with non-negative weights, formally defined as follows: Given 
\iporacle{} oracle access to the adjacency matrix $A$~\footnote{Assume 
that $G$ is a complete graph such that the weights on $\{i,j\} \notin E(G)$ 
is $0$. Also, in the adjacency matrix of $A$, $A_{ij}$ is the weight on 
the edge $\{i,j\}$.} of a graph $G$, the objective is to estimate the quantity $Q=\sum\limits_{\{i,j\} \in E(G)} A_{ij}$. Observe that as $2Q = \sum\limits_{1 \leq i, \, j \leq n} A_{ij}$. From our results on \onebilinear as mentioned in Table~\ref{table:graphresults1-main}, $Q$ can 
be estimated by making $\widetilde{\Oh}\left(\frac{\sqrt{\rho} |V(G)|}{\sqrt{Q}}\right)$ many \iporacle{} queries, where the weights on the edges of $G$ are in $[[\rho]]$.
This is a generalization of the edge estimation results using local queries by Feige~\cite{Feige06}, and Goldreich and Ron~\cite{GoldreichR08}.
Also, according to our results on \onealmostuni, we can design an almost uniform sampler of $E(G)$, i.e., an edge $\{i,j\} \in E(G)$ is sampled with probability $p_{ij}$ satisfying the following inequality:
$$
    (1-\eps)\frac{A_{ij}}{\sum\limits_{\{k,l\} \in E(G)} A_{kl}} \leq p_{ij} \leq (1+\eps)\frac{A_{ij}}{\sum\limits_{\{l,k\} \in E(G)} A_{kl}}.
$$

The sampler also makes $\widetilde{\Oh}\left(\frac{\sqrt{\rho} |V(G)|}{\sqrt{Q}}\right)$ many queries to the \iporacle{} oracle. This is a generalization of a result in the unweighted graph setting by Eden and Rosenbaum~\cite{EdenR18}.

\section{Other Matrix Problems}
\label{app:matrix}

\noindent {Recently, vector-matrix query~\cite{SWYZ19} and vector-matrix-vector query~\cite{DBLP:conf/approx/RashtchianWZ20} were introduced to study a bunch of matrix, graph and statistics problems. As noted earlier,  \iporacle{} query oracle is in the same linear algebraic framework of vector-matrix query and vector-matrix-vector query, but these queries are stronger than \iporacle{} query. Study of the various matrix, graph and statistics problems, introduced in~\cite{SWYZ19,DBLP:conf/approx/RashtchianWZ20}, using \iporacle{} query will be of independent interest. As a first step in that direction, we study the query complexity of the following problems using \iporacle{} queries.}
\begin{itemize}
    \item {\sc Symmetric Matrix}: Is $A\in \{0,1\}^{n\times n}$ a symmetric matrix? 
    
    \item {\sc Diagonal Matrix}: Is $A$ a diagonal matrix? 
    
    \item {\sc Trace}: Compute the trace of the matrix $A$.
    
    \item {\sc Permutation Matrix}: Is $A \in \{0,1\}^{n\times n}$ a permutation matrix?
    
    \item {\sc Doubly stochastic matrix}: Is $A \in \{0,1\}^{n\times n}$ a doubly stochastic matrix?
    
    \item {\sc Identical columns}: Does there exist two columns in $A\in \{0,1\}^{n\times n}$ that are identical?
    
    \item {\sc All ones column}: Does there exist a column in $A$ all of whose entries are $1$?
\end{itemize}

\begin{table}[h]
\centering
\caption{Other matrix problems.}

\begin{tabular}{||c | c |c ||} 
 \hline
 {\bf Problem} & {\bf Query} & {\bf Comments}\\
 \hline\hline
  {\sc Symmetric Matrix} & $\widetilde{\Theta}\left(n\right)$ & Theorem~\ref{theo:ub_symm} \\ 
 \hline
 {\sc  Diagonal Matrix} &$\Theta\left(n\right)$ & Theorem~\ref{theo:ub_dia} \\ \hline
 {\sc Trace } & $\Theta\left(n\right)$& Theorem~\ref{theo:ub_trace} \\ \hline
  {\sc Permutation Matrix} &${\Theta}\left(n\right)$ & Theorem~\ref{theo:ub_perm}\\ \hline
  {\sc Doubly Stochastic matrix } &${\Theta}\left(n\right)$ &Theorem~\ref{theo:ub_sto} \\ \hline
    {\sc  Identical columns} & $\widetilde{\Theta}\left(n\right)$ & Theorem~\ref{theo:ub_identical} \\ \hline
  {\sc All Ones Columns} & ${\Theta}\left(n\right)$ & Theorem~\ref{theo:ub_allone}\\ \hline
\end{tabular}
\label{table:matrix-main-app}

\end{table}

Table~\ref{table:matrix-main-app} gives the query complexity of solving the above matrix problems using IP oracle. {We also do a comparative study of \iporacle{} with respect to these stronger queries.} The details are as follows:

\input{wood.tex}

%% file: wood.tex



\remove{We have proved the results corresponding to the first two rows of Table~\ref{table:graphresults1-main} in Section~\ref{sec:matrix}. We will prove result corresponding to remaining rows in this Section.}

\begin{theo}[{\sc Symmetric Matrix}]\label{theo:ub_symm}
There exists an algorithm, that given an \iporacle{} access to an unknown matrix $A \in \{0,1\}^{n \times n}$, decides whether $A$ is symmetric or not with high probability by using $\Tilde{\Oh}(n)$ many \iporacle{} queries. Also, any algorithm that has \iporacle{} query access to an unknown matrix $A$ and decides whether $A$ is symmetric or not with probability $2/3$, requires $\Omega(n)$ many \iporacle{} queries to $A$.
\end{theo}
\begin{proof}
    First we prove the upper bound result. Let $t=\Theta(\log n)$. Pick $\boldvec{v_1}, \ldots, \boldvec{v_t} \in \{0,1\}^n$ uniformly at random. For each $j\in [n]$ and $k \in [t]$, find  $D_{\boldvec{v_k}}(j) = \dpd{A_{*j}}{\boldvec{v_k}}- \dpd{A_{j*}}{\boldvec{v_k}}~\mbox{mod}~2$. If there exists one $(j,k) \in [n]\times [t]$ such that $D_{\boldvec{v_k}}(j)\neq 0$, then we report $A$ is not a symmetric matrix. Otherwise, we report $A$ is symmetric. Observe that the number of \iporacle{} queries made by the algorithm is $\widetilde{\Oh}(n)$. 
    
    Now we prove the correctness. If $A$ is symmetric, then  $D_{\boldvec{v_k}}(j)= 0$ for each $(j,k) \in [n]\times [t]$ with probability $1$. Now consider when $A$ is not symmetric. Then there exists $j_1\in [n]$ such that the $j_1$-th row of $A$ is not same as that of the $j_1$-th column, that is, $A_{j_1*}\neq A_{*j_1}$. It can be shown that, for any fixed $k\in [t]$,  $\dpd{A_{*j}}{\boldvec{v_k}}~\mbox{mod}~2 = \dpd{A_{j*}}{\boldvec{v_k}} ~\mbox{mod}~2$, that is, $D_{\boldvec{v_k}}(j)= 0$ with probability $1/2$. As we are taking $t=\Theta(\log n)$ many random vectors,  $\boldvec{v_1}, \ldots, \boldvec{v_t} \in \{0,1\}^n$, the probability that there exists $k\in [t]$ such that $D_{\boldvec{v_k}}(j)\neq 0$ is $1-(1/2)^{\Theta{\log n}}$, that is, our algorithm reports that $A$ is symmetric with high probability.
    
    For the lower bound, consider a variation of classical {\sc Disjointness}~\footnote{{\sc Disjointness} is same as $k$-{\sc Intersection} (See Definition~\ref{defi:k-intersection} in Section~\ref{sec:lbd_proofs}) when $k=1$.}  in communication complexity, where Alice is given $\boldvec{x} \in \{0,1\}^n$ and Bob is given $\boldvec{y} \in \{0,1\}^n$. The players want to output $1$ if there is an $i \in [n]$ such that $x_i = y_{i+1} = 1$, where the indices of $x$ and $y$ are modulo $n$. Furthermore, it is promised that either there is a unique $i \in [n]$  such that $x_i = y_{i+1} = 1$ or there is no such $i$. We name the above problem as
{\sc Shifted Disjointness}. It admits a randomized communication complexity of 
$\Omega(n)$. This follows from the randomized communication complexity of standard {\sc Disjointness}.
    
  We prove the the desired lower bound by a reduction from {\sc Shifted Disjointness}. Consider a matrix $A \in \{0,1\}^{n\times n}$ as follows. For each $i,j \in [n]$ with $j\neq i+1$, $A_{ij}=0$. If $x_i = y_{i+1} = 1$, then $A_{i(i+1))} = 1$, and $0$, otherwise. Note that the indices of $x$ and $y$ are modulo $n$. Observe that $A$ is a symmetric matrix if and only if $\boldvec{x}$ and $\boldvec{y}$ do not intersect. 
    
    We will be done by showing that Alice and Bob can determine the answer to any \iporacle{} query, to matrix $A$, with $2$ bits of communication. Let $\dpd{A_{i*}}{\boldvec{v}}$ be an \iporacle{} query. From the construction of matrix $A$, answer to this query depends on $x_i$ and $y_{i+1}$. So, Alice and Bob can determine $\dpd{A_{i*}}{\boldvec{v}}$ with $2$ bits of communication.
    \remove{
    
    For a fixed $j \in [n]$ and $k \in [t]$, consider 
    $D_{\boldvec{v_k}}(j) = \dpd{A_{*j}}{\boldvec{v_k}}- \dpd{A_{j*}}{\boldvec{v_k}}$. If $A_{*j} = A_{j*}$ then $D_{\boldvec{v_k}}(i) = 0$ with probability $1$. If $A_{*j} \neq A_{j*}$, then $D_{\boldvec{v_k}}(i) \neq 0$ with probability at least $1/2$. So, if

    Boost the success probability to $1/n$ and then use the algorithm for all rows and columns $j \in [n]$.}
\end{proof}

\begin{theo}[{\sc Diagonal Matrix}]
\label{theo:ub_dia}
There exists an algorithm, that given an \iporacle{} access to an unknown matrix $A$, decides whether $M$ is diagonal or not deterministically by using ${\Oh}(n)$ many \iporacle{} queries. Also, any algorithm, that has \iporacle{} query access to an unknown matrix $A$ and decides whether $A$ is a diagonal matrix or not with probability $2/3$, requires $\Omega(n)$ many \iporacle{} queries to $A$.
\end{theo}
\begin{proof}
First, let us discuss the upper bound. For each $i\in [n]$, we determine $A_{ii}=\dpd{A_{i*}}{\boldvec{e_i}}$~\footnote{$\boldvec{e_i}$ is the indicator vector for the $i$-th coordinate.} and $\sum\limits_{j=1}^n A_{ij}=\dpd{A_{i*}}{{\bf 1}}$. If there exists an $i \in [n]$ with $A_{ii}\neq \sum\limits_{j=1}^n A_{ij}$, then the algorithm reports $A$ is not a diagonal matrix. Otherwise, the algorithm reports $A$ is a diagonal matrix. Note that the algorithm makes exactly $2n$ many \iporacle{} queries. The correctness of the algorithm follows from its description.

The proof of the lower bound is similar to the proof of the lower bound part of Theorem~\ref{theo:ub_symm}.
\end{proof}



\begin{theo}[{\sc Trace}]
\label{theo:ub_trace}
There exists an algorithm, that given an \iporacle{} access to an unknown matrix $A$, determines the value of the trace of $A$ deterministically by using ${\Oh}(n)$ many \iporacle{} queries. Also, any algorithm, that has \iporacle{} query access to an unknown matrix $A$ and finds the value of the trace of $A$ with probability $2/3$, requires $\Omega(n)$ many \iporacle{} queries to $A$.
\end{theo}
\begin{proof}
    The upper bound, in the above theorem, follows from the fact that the trace of $A$ is $\sum\limits_{i=1}^n A_{ii}=\sum\limits_{i=1}^n \dpd{A_{i*}}{{\bf e_i}}$.
    
    We prove the lower bound by giving a reduction from {\sc Disjointness}. Consider a matrix $A \in \{0,1\}^{n \times n}$ as follows. For each $i,j \in [n]$ with $j\neq i$, $A_{ij}=0$~\footnote{Here $0$ is arbitrary. Any other real number is also ok for our purpose.}. If $x_i = y_{i} = 1$, then $A_{ii} = 1$, and $0$, otherwise. Observe that the trace of $A$ is $1$ if $\boldvec{x}$ and $\boldvec{y}$ intersect. Otherwise, the trace of $A$ is $0$.
    We will be done with the proof by showing that Alice and Bob can determine the answer to each \iporacle{} query with $2$ bits of communication as follows. To determine $\dpd{A_{i*}}{\boldvec{v}}$, Alice and Bob needs to determine $A_{ii}$ that depends on $x_i$ and $y_i$. It is because both Alice and Bob know $A_{ij}=1$ with $j \neq i$.  
\end{proof}


\begin{theo}[{\sc Permutation matrix}]\label{theo:ub_perm}
There exists an algorithm, that given an \iporacle{} access to an unknown matrix $A$, decides whether $A$ is a permutation matrix or not deterministically by using $\Tilde{\Oh}(n)$ many \iporacle{} queries. Also, any algorithm that has \iporacle{} query access to an unknown matrix $A$ and decides whether $A$ is a permutation matrix with probability $2/3$, requires $\Omega(n)$ many \iporacle{} queries to $A$.
\end{theo}
\begin{proof}
A matrix $A$ is a permutation matrix if and only if each row and column of $A$ has exactly one $1$. The upper bound part of the above theorem follows from the following algorithm:
for each $i \in [n]$, we check whether both $\dpd{A_{i*}}{\boldvec{1}} = 1$ and $\dpd{A_{*i}}{\boldvec{1}} = 1$. If for all $i \in [n]$, $\dpd{A_{i*}}{\boldvec{1}} = \dpd{A_{*i}}{\boldvec{1}} = 1$, then $A$ is a permutation matrix. Otherwise, $A$ is not a permutation matrix. The number of queries made by the algorithm is $2n$. The correctness of the algorithm follows from the description of the algorithm.

 For the lower bound, we prove by giving a reduction from {\sc Disjointness}. Consider a matrix $A \in \{0,1\}^{n \times n}$ as follows. For each $i,j \in [n]$ with $j\neq i$, $A_{ij}=0$~\footnote{Here $0$ is arbitrary. We can take any real number here.}. If $x_i = y_{i} = 1$, then $A_{ii} = 0$, and $1$, otherwise. Observe that $A$ is a permutation matrix if and only if $\boldvec{x}$ and $\boldvec{y}$ do not intersect. 
  Observe that Alice and Bob can determine the answer to each \iporacle{} query with $2$ bits of communication. The argument is same as that of in the proof of Theorem~\ref{theo:ub_trace}.
\end{proof}


\begin{theo}[{\sc Doubly stochastic}]
\label{theo:ub_sto}
There exists an algorithm, that given an \iporacle{} access to an unknown matrix $A\in \{0,1\}^{n \times n}$, decides whether $A$ is doubly stochastic or not deterministically by using ${\Oh}(n)$ many \iporacle{} queries. Also, any algorithm that has \iporacle{} query access to an unknown matrix $A$ and decides whether $A$ is a doubly stochastic matrix with probability $2/3$, requires $\Omega(n)$ many \iporacle{} queries to $A$.
\end{theo}
\begin{proof}
A matrix $A$ is a doubly stochastic matrix if and only if the sum of the entries in all rows and columns are same. The upper bound part of the above theorem follows from the following algorithm:
for each $i \in [n]$, determine $\sum_{j=1}^n A_{ij}= \dpd{A_{i*}}{\boldvec{1}}$ and $\sum\limits_{j=1}^n A_{ji}=\dpd{A_{*i}}{\boldvec{1}}$. If for each $i\in [n]$,  $\dpd{A_{i*}}{\boldvec{1}}=\dpd{A_{*i}}{\boldvec{1}}=1$, then the algorithm reports that $A$ is doubly stochastic. Otherwise, the algorithm reports $A$ is not doubly stochastic. The number of queries made by the algorithm is $2n$. The correctness of the algorithm follows from the description of the algorithm.

The proof of the lower bound is similar to that of the proof of the lower bound part of Theorem~\ref{theo:ub_perm}.
\end{proof}


\begin{theo}[{\sc Identical Columns}]\label{theo:ub_identical}
There exists an algorithm, that given an \iporacle{} access to an unknown matrix $A\in \{0,1\}^{n \times n}$, decides whether $A$ has two columns that are identical or not with high probability by using $\Tilde{\Oh}(n)$ many \iporacle{} queries.  Also, any algorithm that has \iporacle{} query access to an unknown matrix $A$ and decides whether $A$ has a pair of identical columns or not with probability $2/3$, requires $\Omega(n)$ many \iporacle{} queries to $A$.
\end{theo}
\begin{proof}
First, we discuss the upper bound.  Let $t=\Theta(\log n)$. Pick $\boldvec{v_1}, \ldots, \boldvec{v_t} \in \{0,1\}^n$ uniformly at random. For each $(r,s)  \in [n] \times [n]$, such that $r \neq s$, and $k \in [t]$, find  $D_{\boldvec{v_k}}(r,s) = \dpd{A_{*r}}{\boldvec{v_k}}- \dpd{A_{*s}}{\boldvec{v_k}}~\mbox{mod}~2$. If there exists $(r,s) \in [n]\times [n]$ such that $D_{\boldvec{v_k}}(r,s)= 0$ for all $k \in [t]$, then we report $A$ is has two identical columns. Otherwise, we report no two columns of $A$ are identical. Observe that the number of \iporacle{} queries made by the algorithm is $\widetilde{\Oh}(n)$. Now we prove the correctness of the above algorithm. If two columns of $A$ are identical, say $r,s \in [n]$, $r\neq s$, then  $D_{\boldvec{v_k}}(r,s) = 0$ for all $k \in [t]$  with probability $1$. Now consider when no two columns of $A$ are identical. Consider a fixed $(r,s) \in [n]\times [n]$ with $r \neq s$
   and $k\in [t]$.  $\dpd{A_{*r}}{\boldvec{v_k}}~\mbox{mod}~2 = \dpd{A_{*s}}{\boldvec{v_k}} ~\mbox{mod}~2$, that is $D_{\boldvec{v_k}}(r,s) = 0$ holds probability $1/2$. As we are taking $t=\Theta(\log n)$ many random vectors,  $\boldvec{v_1}, \ldots, \boldvec{v_t} \in \{0,1\}^n$, the probability that there exists $k\in [t]$, such that $D_{\boldvec{v_k}}(r,s)\neq 0$ is $1-(1/2)^{\Theta{\log n}}$, that is, the algorithm decides that the $r$-th column is different from $s$-th column with probability $1-(1/2)^{\Theta{\log n}}$. Using union bound over all $(r,s) \in [n]\times [n]$, the algorithm reports \emph{no two columns are identical} with probability at least $1-{n^2}(1/2)^{\Theta(\log n)}$.  The algorithm, described above, not only decides whether there exists two identical columns, but also can compute the number of pairs of identical columns with high probability.
   
    We prove the lower bound by giving a reduction from {\sc Disjointness} problem. Consider a matrix $A \in \{0,1\}^{n \times n}$ constructed from ${I} \in \{0,1\}^{n \times n}$ as follows, where $I$ is the $(n \times n)$ identity matrix. If there exists $i \in [n]$ such that $x_i = y_{i} = 1$, then $A$ has the $i$-th and $(i+1)$-th column of ${I}$ changed to $\boldvec{1}$, where the indices $i$ and $(i+1)$ are $\text{mod }n$. Otherwise $A = { I}$. Note that the indices of $\boldvec{x}$ and $\boldvec{y}$ are $\mbox{mod n}$. Observe that $A$ has two identical columns if and only if $\boldvec{x}$ and $\boldvec{y}$ intersect.
    
    We will be done with the proof by showing that Alice and Bob can determine the answer to each \iporacle{} query with $4$ bits of communication as follows. To determine $\dpd{A_{i*}}{\boldvec{v}}$, Alice and Bob needs to determine $A_{i*}$ that depends only on either $x_i$ and $y_{i}$, or $x_{i-1}$ and $y_{i-1}$.
\end{proof}

\begin{theo}[{\sc All One Columns}]
\label{theo:ub_allone}
There exists an algorithm, that given an \iporacle{} access to an unknown matrix $A$, decides whether $A$ has at least one all $1$'s columns or not deterministically by using ${\Oh}(n)$ many \iporacle{} queries. Note that here the entries of matrix $A$ are $0$ or $1$.  Also, any algorithm that has \iporacle{} query access to an unknown matrix $A$ and decides whether $A$ has an all $1$'s column with probability $2/3$, requires $\Omega(n)$ many \iporacle{} queries to $A$.
\end{theo}
\begin{proof}
For each $i \in [n]$, determine $\sum_{j=1}^n A_{ji}= \dpd{A_{*i}}{\boldvec{1}}$. If for each $i\in [n]$ with $\dpd{A_{i*}}{\boldvec{1}}=\dpd{A_{*i}}{\boldvec{1}}=1$, then the algorithm reports that $A$ has at least one all $1$'s column. Otherwise, the algorithm reports $A$ does not have an all $1$'s column. The number of queries made by the algorithm is $n$. The correctness of the algorithm follows from the description of the algorithm.

The lower bound proof is similar to the proof of the lower bound part of 
Theorem~\ref{theo:ub_identical}.
\end{proof}

\subsection*{Comparison with other matrix queries}

\begin{table}[h]
    \caption{Comparing \iporacle{} with matrix-vector and vector-matrix-vector queries.}
    \centering
    \begin{tabular}{||c | c |c |c ||} 
 \hline
 {\bf Problem} & {\bf \iporacle{}} & {\bf $\boldvec{x}^T A \boldvec{y}$} & {\bf $A \boldvec{y}$}\\
 & {\bf Query} &   {\bf Query}~\cite{DBLP:conf/approx/RashtchianWZ20} & {\bf Query}~\cite{SWYZ19}\\
 \hline\hline
  {\sc Symmetric Matrix} & $\widetilde{\Theta}\left(n\right)$ 
  & $\Oh\left(1\right)$ & $\Oh\left(1\right)$ \\ 
  & (Theorem~\ref{theo:ub_symm})
  &  & \\
 \hline
 {\sc  Diagonal Matrix} &$\Theta\left(n\right)$ & $\Oh\left(1\right)$ & $\Oh\left(1\right)$ \\ 
 & (Theorem~\ref{theo:ub_dia})
  &  & \\
 \hline
 {\sc Trace } & $\Theta\left(n\right)$ & $\Oh\left(n\right)$  & $\Oh\left(n\right)$\\
 & (Theorem~\ref{theo:ub_trace})
  &  $\Omega\left( \frac{n}{\log n}\right)$ & $\Omega\left( \frac{n}{\log n}\right)$ \\
 \hline
  {\sc Permutation Matrix} &${\Theta}\left(n\right)$ & $\Oh\left(1\right)$ & $\Oh\left(1\right)$ \\ 
  & (Theorem~\ref{theo:ub_perm})
  &  & \\
  \hline
  {\sc Doubly Stochastic matrix } &${\Theta}\left(n\right)$ & $\Oh\left(1\right)$ & $\Oh\left(1\right)$ \\ 
  & (Theorem~\ref{theo:ub_sto})
  &  & \\
  \hline
    {\sc  Identical columns}\tablefootnote{Upper and lower bounds results for {\sc Identical columns} problem in \cite{DBLP:conf/approx/RashtchianWZ20,SWYZ19} uses vectors from $\{0,1\}^{n}$ in their respective queries.} & $\widetilde{\Theta}\left(n\right)$ & $\widetilde{\Theta}\left(n\right)$  & $\widetilde{\Theta}\left(n\right)$ \\ 
    & (Theorem~\ref{theo:ub_identical})
  &  & \\
    \hline
  {\sc All Ones Columns} & ${\Theta}\left(n\right)$ & $\Oh\left(n\right)$ & $\Oh\left(n\right)$ \\
  & (Theorem~\ref{theo:ub_allone})
  &  $\Omega\left( \frac{n}{\log n}\right)$ & $\Omega\left( \frac{n}{\log n}\right)$ \\
  \hline
\end{tabular}
    \label{tab:comparison_table}
\end{table}

We investigated the problems mentioned in Table~\ref{tab:comparison_table} using \iporacle{}. It will be interesting to study other matrix and graph problems mentioned in~\cite{SWYZ19,DBLP:conf/approx/RashtchianWZ20} using \iporacle{}.

\remove{
\begin{center}
\begin{tabular}{c|c|c}
     Problem & Upper Bound & Lower Bound\\
     \hline
     \symmetric & $\Tilde{\Oh}(n)$ & $\Omega(n)$\\
     \diagonal & $\Oh(n)$ & $\Omega(n)$\\
     \trace & $\Oh(n)$ & $\Omega(n)$\\
     \permutation & $\Oh(n)$ & $\Omega(n)$\\
     \doblestochastic & $\Oh(n)$ & $\Omega(n)$\\
     \allonescolumn & $\Oh(n)$ & $\Omega(n)$\\
     \identicalcolumns & $\Tilde{\Oh}(n)$ & $\Omega(n)$
\end{tabular}   
\end{center}

}

%% file: missingproofs.tex
\section{Communication Complexity}
\label{sec:comm-comp}
\noindent
In two-party communication complexity there are two parties, Alice and Bob, that wish to compute a function $\Pi:\{0,1\}^N \times \{0,1\}^N \to \{0,1\}$. Alice is given ${\bf x} \in \{0,1\}^N$ and Bob is given ${\bf y} \in \{0,1\}^N$. Let $x_i~(y_i)$ denotes the $i$-th bit of ${\bf x}~({\bf y})$. While the parties know the function $\Pi$, Alice 
does not know ${\bf y}$, and similarly Bob does not know ${\bf x}$. Thus they communicate bits following a pre-decided protocol $\cP$ in order to compute $\Pi({\bf x},{\bf y})$. We say a randomized protocol $\cP$ computes $\Pi$ if for all $({\bf x},{\bf y}) \in \{0,1\}^N \times \{0,1\}^N$ we have $\pr[\cP({\bf x},{\bf y}) = \Pi({\bf x},{\bf y})] \geq 2/3$. The model provides the parties access to common random string of arbitrary length. The cost of the protocol $\cP$ is the maximum number of bits communicated, where maximum is over all inputs $({\bf x},{\bf y}) \in \{0,1\}^N \times \{0,1\}^N$. The communication complexity of the function is the cost of the most efficient protocol computing $\Pi$. For more details on communication complexity see~\cite{KN97}.

\input{app-bilin.tex}
\input{extensionsandapplications.tex}

%% file: extensionsandapplications.tex
\label{app:results}

%% file: probabilityresults.tex
\section{Probability Results}
\label{sec:prob}

\begin{lem} [See~\cite{Dubhashi09}]
\label{lem:chernoff}
Let $X = \sum_{i \in [n]} X_i$ where $X_i$, $i \in [n]$, are independent random variables,  $X_i \in [0,1]$ and $\mathbb{E}[X]$ is the expected value of $X$. Then
\begin{itemize}
\item[(i)] \label{ch01} For $\epsilon > 0$,
	$\Pr \left[ \left|X - \mathbb{E}[X] \right| > \epsilon \mathbb{E} \left[ X \right] \right] \leq \exp{ \left(- \frac{\epsilon^2}{3} \mathbb{E}[X]\right)}$.
 
\remove{\item[(ii)] \label{ch02} if $t > 2 \mathbb{E}[X]$, then 
\begin{eqnarray}
	\Pr[ X > t ] \leq 2^{-t}
\end{eqnarray}}
\item[(ii)] \label{ch03} Suppose $\mu_L \leq \mathbb{E}[X] \leq \mu_H$, then for $0 < \epsilon < 1$
\begin{itemize}
\item[(a)] $\Pr[ X > (1+\epsilon)\mu_H] \leq \exp{\left( - \frac{\epsilon^2}{3} \mu_H\right)}$.
\item[(b)] $\Pr[ X < (1-\epsilon)\mu_L] \leq \exp{\left( - \frac{\epsilon^2}{2} \mu_L\right)}$.
\end{itemize}
\end{itemize}
\end{lem}